\documentclass[paper,notoc]{JHEP3}

\usepackage{epsfig}
\usepackage{graphicx}
\usepackage{feynmp}
\usepackage{amsmath}

\newcommand{\beq}{\begin{equation}}
\newcommand{\eeq}{\end{equation}}

\def\bsp#1\esp{\begin{split}#1\end{split}}

\newcommand{\bea}{\begin{eqnarray}}
\newcommand{\eea}{\end{eqnarray}}
\newcommand{\nn}{\nonumber}

\def\eqn#1{Eq.~(\ref{#1})}
\def\eqns#1#2{Eqs.~(\ref{#1}) and~(\ref{#2})}

\def\fig#1{Fig.~{\ref{#1}}}
\def\sec#1{Section~{\ref{#1}}}

\newcommand\fverb{\setbox\pippobox=\hbox\bgroup\verb}
\newcommand\fverbdo{\egroup\medskip\noindent%
                        \fbox{\unhbox\pippobox}\ }
\newcommand\fverbit{\egroup\item[\fbox{\unhbox\pippobox}]}
\newbox\pippobox

\def\ord{{\cal O} }
\def\cM{{\cal M}}

\def\eps{\epsilon}

\def\cg{c_\Gamma}
\newcommand\sss{\scriptscriptstyle}
\newcommand\as{\alpha_{\sss S}} 
\newcommand\gs{g}

\def\tgs{\bar\gs}
\def\bC{\bar C}
\def\bV{\bar V}
\def\bW{\bar W}
\def\bA{{\bar A}}
\def\balpha{\bar\alpha}

\newcommand{\ie}{\emph{i.e.}~}

\newcommand{\li}{\textrm{Li}_2}
\newcommand{\im}{\textrm{Im}}

\title{Iterated amplitudes in the high-energy limit}

\author{Vittorio Del Duca\\
Istituto Nazionale di Fisica Nucleare\\
Laboratori Nazionali di Frascati\\
00044 Frascati (Roma), Italy\\
        E-mail: \email{delduca@lnf.infn.it}}

\author{Claude Duhr\\
Institut de Physique Th\'eorique \&
Centre for Particle Physics and Phenomenology (CP3)\\
Universit\'e Catholique de Louvain\\
Chemin du Cyclotron 2,
B-1348 Louvain-la-Neuve, Belgium\\
E-mail: \email{claude.duhr@uclouvain.be}}

\author{E.~W.~N.~Glover\\
Institute for Particle Physics Phenomenology, 
University of Durham\\ Durham, DH1 3LE, U.K.\\
E-mail: \email{E.W.N.Glover@durham.ac.uk}}

\abstract{We consider the high-energy limits of the colour ordered
four-, five- and six-gluon
MHV amplitudes of the maximally supersymmetric QCD in the 
multi-Regge kinematics
where all the gluons are strongly ordered in rapidity.
We show that various building
blocks occurring in the Regge factorisation (the Regge trajectory, the
coefficient functions and the Lipatov vertex) satisfy an iterative structure
very similar to the Bern-Dixon-Smirnov (BDS) ansatz. This iterative structure,
combined with the universality of the building blocks,  enables us to show that
in the Euclidean region
any two- and three-loop amplitude in multi-Regge kinematics  is guaranteed to
satisfy the BDS ansatz. 
We also consider slightly more general  kinematics where
the strong rapidity ordering applies to all the gluons except the two with
either the largest or smallest rapidities, and we derive the iterative formula
for the associated coefficient function. We show that in this kinematic limit
the BDS ansatz is also satisfied. Finally, we argue that only for more general
kinematics - e.g. with three gluons having similar rapidities, or where the two
central gluons have similar rapidities - can
a disagreement with the BDS ansatz arise.}

\preprint{~IPPP/08/63\\
~CP3-08-40}

\begin{document}

\section{Introduction}
\label{sec:intro}

Recently, Bern, Dixon and Smirnov have proposed an ansatz~\cite{Bern:2005iz} 
for the colour-stripped $l$-loop $n$-gluon scattering amplitude in the maximally supersymmetric 
$\begin{cal}N\end{cal}=4$ Yang-Mills theory (MSYM), with the maximally-helicity violating (MHV) configuration
for arbitrary $l$ and $n$.
They checked that the ansatz agrees analytically with the evaluation 
of the three-loop four-gluon amplitude. 
The ansatz has been proven to be correct also for the two-loop five-gluon amplitude,
which has been computed numerically~\cite{Bern:2006vw,Cachazo:2008vp}.
The ansatz implies a tower of iteration formul\ae, which allow one to determine the $n$-gluon
amplitude at a given number of loops in terms of amplitudes with fewer loops.
For example, the iteration formula for the colour-stripped two-loop MHV amplitude $m_n^{(2)}(\eps)$ is
\beq
m_n^{(2)}(\eps) = \frac{1}{2} \left[m_n^{(1)}(\eps)\right]^2
+ f^{(2)}(\eps)\, m_n^{(1)}(2\eps) + Const^{(2)} + \ord(\eps)\, ,\label{eq:ite2bds}
\eeq
thus the two-loop amplitude is determined in terms of a constant, $Const^{(2)}$, 
a known function, $f^{(2)}(\eps)$, of the
dimensional-regularisation parameter $\eps$ (which is related to the 
cusp~\cite{Korchemsky:1987wg,Beisert:2006ez} and 
collinear~\cite{Magnea:1990zb,Sterman:2002qn} anomalous dimensions)
and the one-loop MHV amplitude $m_n^{(1)}(\eps)$ evaluated to $\ord(\eps^2)$.

The BDS ansatz was first predicted to fail by Alday and Maldacena~\cite{Alday:2007he,Alday:2007hr},
for amplitudes with a large number of gluons in the strong-coupling limit.
They claimed that the finite pieces of the two-loop amplitudes with six or more gluons
would be incorrectly determined. One can characterise this statement by the quantity $R_n^{(2)}$
\beq
R_n^{(2)} = m_n^{(2)}(\eps) - \frac{1}{2} \left[m_n^{(1)}(\eps)\right]^2
- f^{(2)}(\eps)\, m_n^{(1)}(2\eps) - Const^{(2)}\, ,\label{eq:discr}
\eeq
where $R_n^{(2)}$ may be a function of the kinematical parameters of the $n$-gluon amplitude,
but a constant with respect to $\eps$. Then the claim was that $R_n^{(2)}\ne 0$ for $n\ge 6$.
This prediction was backed up by Drummond et al.~\cite{Drummond:2007bm}, who
considered the finite contribution to the
hexagonal light-like Wilson loop at two loops. The conclusion was that either the BDS ansatz
was wrong, or the equivalence between Wilson loops and scattering amplitudes did not work at 
two loops. The question was settled in Ref.~\cite{Bern:2008ap,Cachazo:2008hp} 
by the numerical calculation of $m_6^{(1)}(\eps)$ to $\ord(\eps^2)$ and 
of $m_6^{(2)}(\eps)$, which allowed for the
numerical evaluation of $R_6^{(2)}$ and showed that it was different from zero. 
This result also confirmed the equivalence between the scattering amplitude and the finite part of 
the light-like hexagon Wilson loop~\cite{Drummond:2008aq}.

The question remains of how one can determine the function $R_n^{(2)}$?  A direct analytical evaluation in general
kinematics is currently beyond our capability: it would require the computation of  the one-loop hexagon to
$\ord(\eps^2)$,  as well as the two-loop hexagon through to $\ord(\eps^0)$.   Another approach is to try to 
constrain  $R_n^{(2)}$ using some simplified kinematics, where one knows that the amplitude has certain
factorisation properties.  Examples include the limit where one or more of the gluons 
are soft or where two or more of the
gluons are collinear.  In this paper, we consider another limit where the kinematics is simplified - the high
energy limit (HEL).  For a multiparticle process there are several high energy limits that one can take,
corresponding to two or more of the gluon rapidities being strongly ordered, together with constraints on the
transverse momenta of the gluons. By relaxing the restriction on the gluon rapidities and transverse momenta, one
can systematically return to the general kinematics.    
The simplest kinematics corresponds
to the multi-Regge kinematics~\cite{Kuraev:1976ge}, where all of the produced gluons are
strongly ordered in rapidity and have comparable transverse momenta.   
We shall start then with the simplest possible kinematics and we will show that, in the Euclidean region,
$R_n^{(2)}$ does not contribute for any $n$.
Then we shall consider various quasi-multi-Regge kinematics, which gradually approach the more general kinematics, with a
view to determining where the function $R_n^{(2)}$ might not vanish and could therefore be constrained by the HEL.

Our paper is organised as follows.   In Section~\ref{sec:mrkforall}, we review the multi-Regge kinematics and
discuss the Regge factorisation that tree-level (colour stripped) amplitudes obey. In Section~\ref{sec:npthel}, 
we extend the Regge factorisation beyond leading order and provide a conjecture  for the factorised form for the
colour stripped  $n$-gluon amplitude to all orders, both in the Euclidean region, where all invariants are
space-like, and in the physical region, where the $s$-type invariants are time-like and the $t$-type invariants are 
space-like\footnote{For the one-loop six-gluon amplitude, 
in the Minkowski region where the centre-of-mass energy squared $s$ and the energy squared $s_2$ 
of the two gluons emitted along the ladder are time-like while all other invariants stay space-like, the factorised form
conjectured in Section~\ref{sec:npthel} is not valid~\cite{Bartels:2008ce, Schabinger:2009bb}.}.
The high-energy limits of the four-, five- and six-gluon MHV amplitudes are
developed  in section~\ref{sec:456pthel}, including explicit expressions for the Regge trajectory (up to three
loops),  the coefficient functions (up to three loops)  and the Lipatov vertex in MSYM.  In
section~\ref{sec:bdsmrk} we consider the BDS ansatz in the multi-Regge kinematics.  By considering the four- and
five-point amplitudes, we show that both the coefficient function and the Lipatov vertex satisfy an iterative
structure very similar to the BDS ansatz 
itself\footnote{It is well known that the $l$-loop Regge trajectory is
directly related to $f^{(l)}(\eps)$.}. This iterative structure ensures that the six-point amplitude is completely
determined by known functions, and, in the multi-Regge kinematics is guaranteed to satisfy the BDS ansatz in the Euclidean and in the
physical regions. In other words, in those regions the remainder function $R_6^{(2)}$ vanishes in the multi-Regge kinematics. 
We derive exponentiated forms for the
coefficient functions and Lipatov vertex in section~\ref{sec:proof} and prove that we recover 
the BDS ansatz in the multi-Regge kinematics for any number of loops.
We consider other quasi-multi-Regge kinematics in
section~\ref{sec:quasi}. In particular, we consider the slightly more general  kinematics where all but two of the
gluons (the two gluons with either the largest or smallest rapidities) are strongly ordered in rapidity.   This
quasi-multi-Regge kinematics first occurs in the five gluon amplitude and introduces a new coefficient function
with two final state gluons which also satisfies an iterative structure similar to the BDS ansatz.  Once again,
$R_6^{(2)}$ does not contribute in this limit and we note that the   conformal kinematic ratios also take a
particularly simple form in this quasi-multi-Regge kinematics. Finally, in \sec{sec:outlook} we consider
more general kinematics - with
three gluons having similar rapidities, or where the two central gluons have similar rapidities.   These
configurations first appear with four gluons in the final state.   The new vertices and coefficient functions
associated with these kinematics  cannot be determined using the five-gluon amplitude, and require explicit
knowledge of the six-gluon amplitude.   We therefore cannot say anything about the sensitivity of the HEL to
$R_6^{(2)}$, but note that in each of these cases, the three
conformal kinematic ratios relevant for six-gluon scattering do not simplify, and take general
values. We enclose appendices detailing the multi-Regge and quasi-multi-Regge kinematics.

\section{Multi-Regge kinematics}
\label{sec:mrkforall}

Because in this work we make repeated use of the multi-Regge kinematics,
we shall give here a short pedagogical introduction to it. We consider an
$n$-gluon amplitude, $g_1\,g_2\to g_3\,g_4\,\cdots g_n$, with
all the momenta taken as outgoing,
and label the gluons cyclically clockwise. In the 
multi-Regge kinematics~\cite{Kuraev:1976ge}, the produced gluons are
strongly ordered in rapidity and have comparable transverse momenta,
\begin{equation}
y_3 \gg y_4\gg \cdots\gg y_n;\qquad |p_{3\perp}| \simeq |p_{4\perp}| ...\simeq|p_{n\perp}|\, 
.\label{mrknpt}
\end{equation}
Accordingly, we can write the Mandelstam invariants in the approximate 
form\footnote{In Appendices~\ref{sec:mpk} and \ref{sec:mrk}, we write the invariants (\ref{invb}) 
and the spinor products (\ref{ypro}), in terms of light-cone coordinates. Although the light-cone
formulation is more convenient for performing calculations, we prefer to give here
those quantities in terms of rapidities because it is physically more intuitive.
The translation between light-cone coordinates and rapidities is straightforward 
(please see Appendix~\ref{sec:mpk}).}
\begin{eqnarray}
s_{12} &\simeq& |p_{3\perp}| |p_{n\perp}| e^{y_3-y_n}\, ,\nonumber \\
s_{2i} &\simeq& - |p_{3\perp}| |p_{i\perp}| e^{y_3-y_i}\, ,\label{invb}\\
s_{1i} &\simeq& - |p_{i\perp}| |p_{n\perp}| e^{y_i-y_n}\, ,\nonumber\\
s_{ij} &\simeq& |p_{i\perp}| |p_{j\perp}| e^{|y_i-y_j|}\, .\nonumber
\end{eqnarray}
for $i,j = 3,\ldots ,n$. We label the momenta transferred in the $t$-channel as
\bea
q_1 &=& p_1+p_n \nn\\
q_2 &=& q_1+p_{n-1} = q_3 - p_{n-2} \nn\\
&\vdots& \label{eq:mome}\\
q_{n-4} &=& q_{n-5} + p_5 = q_{n-3} - p_4 \nn\\
q_{n-3} &=& -p_2-p_3\, ,\nn
\eea
with virtualities $t_i = q_i^2$.
Then it is easy to see that in the multi-Regge kinematics the transverse
components of the momenta $q_i$ dominate over the longitudinal components,
$q_i^2 \simeq - |q_{i\perp}|^2$. In addition, $t_1=s_{1n}$ and
$t_{n-3}=s_{23}$, and we label $s = s_{12}$, and $s_1=s_{n-1,n}$,
$s_2=s_{n-2,n-1}$, \ldots, $s_{n-3}=s_{34}$ for $n > 4$.
Thus, the multi-Regge kinematics (\ref{mrknpt}) become
\begin{equation}
s \gg s_{1},\ s_{2}, \ldots, s_{n-3} \gg -t_1,\ -t_2,\ldots, -t_{n-3}\, ,\label{eq:mrknpt2} 
\end{equation}
with the special case $s \gg -t$ for $n = 4$.
Labelling the transverse momenta of the gluons emitted along the ladder as
$\kappa_1=|p_{n-1\perp}|^2$, $\kappa_2=|p_{n-2\perp}|^2$, 
\ldots, $\kappa_{n-4}=|p_{4\perp}|^2$, and using \eqn{invb}, we can write
\begin{equation}
\kappa_1 = \frac{s_{1}\, s_{2}}{s_{n-2,n-1,n}} \qquad 
\kappa_2 = \frac{s_{2}\, s_{3}}{s_{n-3,n-2,n-1}} \qquad \cdots  \qquad 
\kappa_{n-4} = \frac{s_{n-4}\, s_{n-3}}{s_{345}}
,\label{massnpt}
\end{equation}
for $n > 4$, which are known as the mass-shell conditions (\ref{eq:masshell})
for the gluons along the ladder. \eqn{invb} also implies a relation amongst the mass-shell
conditions,
\beq
s\, \kappa_1 \cdots \kappa_{n-4} = s_1\, s_2 \cdots s_{n-3}\, .\label{eq:condit}
\eeq

In the multi-Regge kinematics the spinor products are given by \eqn{mrpro}
\begin{eqnarray}
\langle 2 1\rangle &\simeq& -\sqrt{|p_{3\perp}| 
|p_{n\perp}|} \exp\left(\frac{y_3-y_n}{2}\right)\, ,\nonumber\\
\langle 2 i\rangle &\simeq& -i \sqrt{\frac{|p_{3\perp}|}{ |p_{i\perp}|}}\,
p_{i\perp} \exp\left(\frac{y_3-y_i}{ 2}\right)\, ,\label{ypro}\\
\langle i 1\rangle &\simeq& i \sqrt{|p_{i\perp}||p_{n\perp}|}\, \exp\left(\frac{y_i-y_n}{ 2}\right)\, ,\nn\\
\langle i j\rangle &\simeq& -\sqrt{\frac{|p_{i\perp}|}{ |p_{j\perp}|}}\,
p_{j\perp} \exp\left(\frac{y_i-y_j}{2}\right)\, \qquad {\rm for}\, y_i>y_j .\nonumber
\end{eqnarray}

\subsection{MHV amplitudes in multi-Regge kinematics}
\label{sec:mhvmrk}

The colour decomposition of the tree-level $n$-gluon amplitude
is~\cite{Mangano:1990by}
\begin{equation}
\cM_n^{(0)} = 2^{n/2}\, g^{n-2}\, \sum_{S_n/Z_n} {\rm tr}(T^{d_1}
\cdots
T^{d_n}) \, m_n^{(0)}(1,\ldots, n)\, ,\label{one}
\end{equation}
where $d_i$ is the colour of a gluon of momentum $p_i$ and helicity $\nu_i$.
The $T$'s are the colour 
matrices\footnote{We use the normalization
${\rm tr}(T^c T^d) = \delta^{cd}/2$,
although it is immaterial in what follows.} in the
fundamental representation of SU($N$) and the sum is over the noncyclic
permutations $S_n/Z_n$ of the set $[1, \ldots ,n]$. We consider the MHV
configurations $(-,-,+, \ldots ,+)$ for which the tree-level gauge-invariant 
colour-stripped amplitudes  
assume the form
\begin{equation}
m_n^{(0)}(1,2, \ldots ,n) = \frac{\langle p_i p_j\rangle^4}
{\langle p_1 p_2\rangle \cdots\langle p_{n-1} p_n\rangle 
\langle p_n p_1\rangle}\, ,\label{two}
\end{equation}
where $i$ and $j$ are the gluons of negative helicity. The colour structure
of \eqn{one} in multi-Regge kinematics is 
known~\cite{DelDuca:1993pp,DelDuca:1995zy,DelDuca:1999rs}
and will not be considered further.
Here we shall concentrate on the behaviour of the colour-stripped 
amplitudes (\ref{two}), which in multi-Regge kinematics 
has the factorised form~\cite{DelDuca:1995zy}
\begin{eqnarray}
m_n^{(0)}(1,2, \ldots ,n) &=& 
s \left[g\, C^{(0)}(p_2,p_3) \right]\, 
\frac{1}{t_{n-3}}\, \left[g\,V^{(0)}(q_{n-3},q_{n-4};\kappa_{n-4})\right]
\label{treenpt}\\ & &\quad \cdots \times\ \frac{1}{t_2}\,
 \left[g\,V^{(0)}(q_2,q_1,\kappa_1)\right]\, \frac{1}{ t_1}\, 
 \left[g\, C^{(0)}(p_1,p_n) \right]\, .\nonumber
\end{eqnarray}
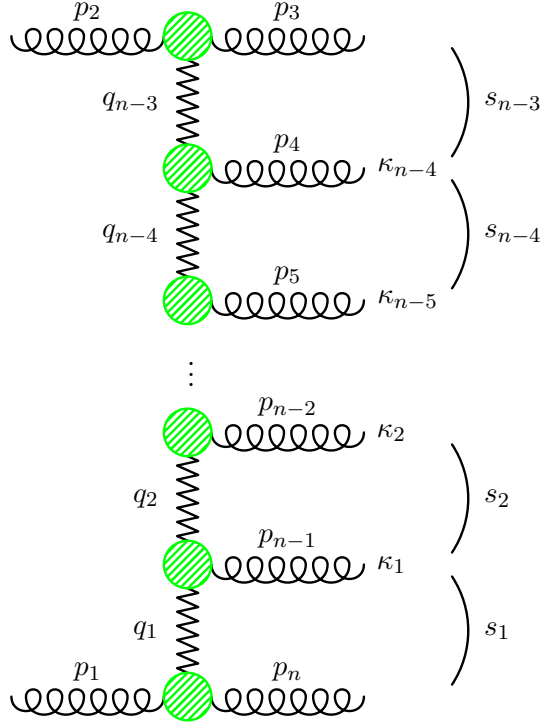
\begin{figure}[!t]
\begin{center}
  \begin{fmffile}{mr}
                  \begin{fmfgraph*}(200,250)
                  \fmfstraight
                     \fmfleft{p1,pdn1,pdn2,pd5,pd4,p2}
                     \fmfright{x1,x2,x3,x5,x6,x7}
                     \fmf{phantom}{p1,u1,v1n,u2,pn,u3,x1}
                     \fmf{phantom}{p2,o1,v23,o2,p3,o3,x7}
                     \fmffreeze
                     \fmf{phantom}{pn,pn1,pn2,p5,p4,p3}
                     \fmffreeze
                     \fmf{gluon,label=$p_2$,label.side=left,l.d=0.03w}{p2,v23}
                     \fmf{gluon,label=$p_3$,label.side=left,l.d=0.03w}{v23,p3}
                     \fmf{phantom}{v23,v4,v5,vn2,vn1,v1n}
                     \fmffreeze
                     \fmfv{decor.shape=circle,decor.filled=shaded,decor.size=.09w,fore=green}{v23}
                     \fmfv{decor.shape=circle,decor.filled=shaded,decor.size=.09w,fore=green}{v4}
                     \fmfv{decor.shape=circle,decor.filled=shaded,decor.size=.09w,fore=green}{v5}
                     \fmfv{decor.shape=circle,decor.filled=shaded,decor.size=.09w,fore=green}{vn2}
                     \fmfv{decor.shape=circle,decor.filled=shaded,decor.size=.09w,fore=green}{vn1}
                     \fmfv{decor.shape=circle,decor.filled=shaded,decor.size=.09w,fore=green}{v1n}
                     \fmf{zigzag,label=$q_{n-3}$,label.side=right,l.d=0.055w}{v23,v4}
                     \fmf{zigzag,label=$q_{n-4}$,label.side=right,l.d=0.055w}{v4,v5}
                     \fmf{zigzag,label=$q_{2}$,label.side=right,l.d=0.055w}{vn2,vn1}
                     \fmf{zigzag,label=$q_{1}$,label.side=right,l.d=0.055w}{vn1,v1n}
                     \fmf{gluon,label=$p_1$,label.side=left,l.d=0.03w}{p1,v1n}
                     \fmf{gluon,label=$p_{n}$,label.side=left,l.d=0.03w}{v1n,pn}
                     \fmffreeze
                     \fmf{gluon,label=$p_4$,label.side=left,l.d=0.03w}{v4,p4}
                     \fmf{gluon,label=$p_5$,label.side=left,l.d=0.03w}{v5,p5}
                     \fmf{gluon,label=$p_{n-2}$,label.side=left,l.d=0.03w}{vn2,pn2}
                     \fmf{gluon,label=$p_{n-1}$,label.side=left,l.d=0.03w}{vn1,pn1}
                     \fmffreeze
                     \fmf{phantom}{u3,ou4,ou5,oun2,oun1,o3}
                     \fmffreeze
                     \fmf{phantom}{o3,ox1,ox2,ox3,ox4,ox5,ox6,ox7,ox8,ox9,ox10,oun1}
                     \fmffreeze
                     \fmf{plain,tension=0.2,left=0.3,label=$s_{n-3}$}{ox1,ox10}
                     \fmffreeze
                     \fmf{phantom}{oun1,oun1x1,oun1x2,oun1x3,oun1x4,oun1x5,oun1x6,oun1x7,oun1x8,oun1x9,oun1x10,oun2}
                     \fmffreeze
                     \fmf{plain,tension=0.2,left=0.3,label=$s_{n-4}$}{oun1x1,oun1x10}
                     \fmffreeze
                     \fmf{phantom}{ou5,ou5x1,ou5x2,ou5x3,ou5x4,ou5x5,ou5x6,ou5x7,ou5x8,ou5x9,ou5x10,ou4}
                     \fmffreeze
                     \fmf{plain,tension=0.2,left=0.3,label=$s_2$}{ou5x1,ou5x10}
                     \fmffreeze
                     \fmf{phantom}{ou4,ou4x1,ou4x2,ou4x3,ou4x4,ou4x5,ou4x6,ou4x7,ou4x8,ou4x9,ou4x10,u3}
                     \fmffreeze
                     \fmf{plain,tension=0.2,left=0.3,label=$s_1$}{ou4x1,ou4x10}
                     \fmffreeze
                     \fmfv{label=$\kappa_{n-4}$,l.a=-180,l.d=-0.14w}{p4}
                     \fmfv{label=$\kappa_{n-5}$,l.a=-180,l.d=-0.14w}{p5}
                     \fmfv{label=$\kappa_{2}$,l.a=-180,l.d=-0.08w}{pn2}
                     \fmfv{label=$\kappa_{1}$,l.a=-180,l.d=-0.08w}{pn1}
                     \fmf{phantom}{v5,v67,vn2}
                     \fmfv{label=$\vdots$,l.d=-1mm}{v67}
                           \end{fmfgraph*}
                             \end{fmffile}
                             \end{center}
\caption{\label{fig:MR}Amplitude in the multi-Regge kinematics. 
The green blobs indicate the coefficient functions (impact factors) and the Lipatov vertices describing the emission of gluons along the ladder.}
\end{figure}
This factorization is shown schematically in~\fig{fig:MR}.
The gluon coefficient functions $C^{(0)}$, which yield the LO gluon impact factors, 
are given in Ref.~\cite{Kuraev:1976ge} in terms of their spin structure 
and in Ref.~\cite{DelDuca:1995zy,DelDuca:1996km} at fixed
helicities of the external gluons,
\begin{equation}
C^{(0)}(p_2^-,p_3^+) = 1 \qquad C^{(0)}(p_1^-,p_n^+) = \frac{p_{n\perp}^*} 
{p_{n\perp}}\, ,\label{centrc}
\end{equation}
with $p_{\perp}=p_x+ip_y$ the complex transverse momentum.
The vertex for the emission of a gluon along the ladder 
is the Lipatov vertex~\cite{DelDuca:1995zy,Lipatov:1976zz,Lipatov:1991nf}
\begin{equation}
V^{(0)}(q_{j+1},q_{j},\kappa_j) = \sqrt{2}\, \frac{q^*_{{j+1}\perp} q_{{j}\perp}}{p_{n-j\perp}}\, 
,\label{lipeq}
\end{equation}
with $p_{n-j} = q_{j+1} - q_j$.

\section{The high-energy limit of  the $n$-gluon amplitude}
\label{sec:npthel}

The virtual radiative corrections to Eq.~(\ref{treenpt}) in the
leading logarithmic (LL) approximation are obtained, to all orders
in $\as$, by replacing the propagator of the $t$-channel gluon by its
reggeised form~\cite{Kuraev:1976ge}. That is, by making the replacement
\begin{equation}
\frac{1}{ t_i} \to \frac{1}{t_i} 
\left(\frac{s_i}{ \tau}\right)^{\alpha(t_i)}\, ,\label{sud}
\end{equation}
in Eq.~(\ref{treenpt}), where $\alpha(t_i)$ can be written in
dimensional regularization in $d=4-2\epsilon$ dimensions as
\begin{equation}
\alpha(t_i) = \gs^2\, c_{\Gamma}\,  
\left(\frac{\mu^2}{ -t_i}\right)^{\epsilon} \, N\, \frac{2}{\epsilon}
,\label{alph}
\end{equation}
with $N$ colours, and
\begin{equation}
c_{\Gamma} = \frac{1}{(4\pi)^{2-\epsilon}}\, \frac{\Gamma(1+\epsilon)\,
\Gamma^2(1-\epsilon)}{ \Gamma(1-2\epsilon)}\, .\label{cgam}
\end{equation}
$\alpha(t_i)$ is the Regge trajectory and accounts for the
higher order corrections to gluon exchange in the $t_i$ channel. In \eqn{sud},
the reggeisation scale $\tau$ is introduced to separate contributions to
the 
reggeized propagator, the coefficient
function and the Lipatov vertex.  It is much smaller than any of the $s$-type
invariants, and it is of the order of the $t$-type invariants.
In order to go beyond the LL approximation and to compute the higher-order
corrections to the Lipatov vertex (\ref{lipeq}), we need a high-energy
prescription~\cite{Fadin:1993wh} that 
disentangles the virtual corrections to the Lipatov vertex
from those to the coefficient functions (\ref{centrc})
and from those that reggeize the gluon (\ref{sud}). 
The high-energy prescription of Ref.~\cite{Fadin:1993wh} is given 
at the colour-dressed amplitude level in QCD, where it holds
to the next-to-leading-logarithmic (NLL) accuracy. However, it has been
shown to break down in the imaginary part of the QCD one-loop four-parton
amplitude~\cite{DelDuca:1998kx}, in the imaginary part of the QCD
one-loop five-gluon amplitude~\cite{DelDuca:1998cx},
and in the two-loop four-point amplitude in MSYM~\cite{DelDuca:2008pj}.
This is because the mismatches between the colour orderings and the
multi-Regge kinematics become apparent at NLL. When the colour ordering
is correctly aligned with the multi-Regge limit, the factorisation applies to NLL
and beyond. In Ref.~\cite{DelDuca:2008pj}, we showed that the
high-energy prescription, applied to the colour-stripped four-point
amplitude is valid up to three loops.
Thus, we conjecture that in the multi-Regge kinematics in the Euclidean region
a generic colour-stripped $n$-gluon amplitude has the factorised form,
\bea
\lefteqn{ m_n(1,2, \ldots ,n) =
s \left[g\, C(p_2,p_3) \right]\, 
\frac{1}{t_{n-3}}\, \left(\frac{-s_{n-3}}{ \tau}\right)^{\alpha(t_{n-3})}
\left[g\,V(q_{n-3},q_{n-4},\kappa_{n-4})\right]  }
\nn\\ &&\qquad\qquad \cdots \times\ 
{1\over t_2}\, \left({-s_2\over \tau}\right)^{\alpha(t_2)}
\left[g\,V(q_2,q_1,\kappa_1)\right]\, 
{1\over t_1}\, \left({-s_1\over \tau}\right)^{\alpha(t_1)}
 \left[g\, C(p_1,p_n) \right]\, , \label{loopnpt}
\eea
where we suppressed the dependence of the
coefficient function and of the Lipatov vertex on the reggeisation scale $\tau$,
and on the dimensional regularisation parameters $\mu^2$ and $\eps$.

In the Euclidean region, where the invariants are all negative,
\beq
s, s_1, s_2, \ldots, s_{n-3}, t_1, t_2, \ldots t_{n-3} < 0\, ,\label{eq:unphys}
\eeq
the colour-stripped amplitude $m_n$, \eqn{loopnpt}, is real.
Then the multi-Regge kinematics~(\ref{eq:mrknpt2}) are
\begin{equation}
-s \gg -s_{1}, -s_{2}, \ldots, -s_{n-3}\gg -t_1, -t_2 \dots, -t_{n-3}\, ,\label{eq:mrkneg} 
\end{equation}
and the on-shell condition (\ref{massnpt}) is
\begin{equation}
-\kappa_1 = {(-s_{1})\, (-s_{2})\over -s_{n-2,n-1,n}}\, ,\quad 
-\kappa_2 = {(-s_{2})\, (-s_{3})\over -s_{n-3,n-2,n-1}}\, ,\quad \cdots  \quad 
-\kappa_{n-4} = {(-s_{n-4})\, (-s_{n-3})\over -s_{345}}
.\label{nmassnpt}
\end{equation}
In \eqn{loopnpt}, the Regge trajectory has the perturbative expansion,
\begin{equation}
\alpha(t_i) = \tgs^{2} \balpha^{(1)}(t_i) + \tgs^4 \balpha^{(2)}(t_i) + 
\tgs^6 \balpha^{(3)}(t_i) + \ord(\tgs^8)\, ,\label{alphb}
\end{equation}
with $i=1,\ldots,n-3$, and with the rescaled coupling
\beq
\tgs^2 = \gs^2 \cg N\, .\label{rescal}
\eeq
In \eqn{loopnpt}, the coefficient functions $C$ and the Lipatov vertex $V$ are also
expanded in the rescaled coupling,
\begin{eqnarray} 
C(p_i,p_j,\tau) &=& C^{(0)}(p_i,p_j)\left(1 + \sum_{r=1}^{s-1} \tgs^{2r} \bC^{(r)}(t_k,\tau) 
+ \ord(\tgs^{2s}) \right)\, 
,\label{fullv} \\
V(q_{j+1},q_j,\kappa_j,\tau) &=& V^{(0)}(q_{j+1},q_j)\left(1 + \sum_{r=1}^{s-1} \tgs^{2r} 
\bV^{(r)}(t_{j+1},t_j,\kappa_j,\tau)
+ \ord(\tgs^{2s}) \right)\, .\nn
\end{eqnarray}
with $(p_i+p_j)^2=t_k$
where $C$ and $V$ are real,
up to overall complex phases in $C^{(0)}$, \eqn{centrc}, and $V^{(0)}$, 
\eqn{lipeq}, induced by the complex-valued helicity bases.
Note that because several transverse scales 
occur, we prefer to keep the dependence on $\mu^2$ of the trajectory, coefficient
function and Lipatov vertex within the loop coefficient rather than in the 
rescaled coupling,
\bea
&& \balpha^{(n)}(t_i) = \left({\mu^2\over -t_i}\right)^{n\eps} \alpha^{(n)}\, 
,\quad \bC^{(n)}(t_k,\tau) = \left({\mu^2\over -t_k}\right)^{n\eps} 
C^{(n)}(t_k,\tau)\, ,\nn\\
&& \bV^{(n)}(t_{j+1},t_j,\kappa_j,\tau) = \left({\mu^2\over -\kappa_j}\right)^{n\eps} 
V^{(n)}(t_{j+1},t_j,\kappa_j,\tau)\, .\label{eq:coeffrescal}
\eea
The expansion of \eqn{loopnpt} can be written as
\beq
m_n = m_n^{(0)} \left( 1 + \tgs^2\ m_n^{(1)} + \tgs^4 m_n^{(2)} + \tgs^6 m_n^{(3)} 
+ \ord(\tgs^8) \right)\, .\label{elasexpand}
\eeq

\subsection{Analytic continuation of the $n$-gluon amplitude to the physical region}
\label{sec:analytic}

We analytically continue the high-energy prescription for the 
colour-stripped amplitude (\ref{loopnpt}) to the physical region\footnote{Care must be
exercised in analytically continuing \eqn{loopnpt}: in Ref.~\cite{Bartels:2008ce} it has
been shown that in the Minkowski region where $s, s_2$ are positive while all other 
invariants stay negative, the one-loop six-gluon amplitude cannot be cast in the form
of \eqn{loopnpt}.}, where
\beq
s, s_1, s_2, \ldots s_{n-3} > 0\, ,\qquad t_1, t_2, \ldots t_{n-3} < 0\, ,\label{eq:phys}
\eeq
through the usual prescription $\ln(-s_j) = \ln(s_j) - i\pi$, for $s_j > 0$.
Then the multi-Regge kinematics are given by \eqn{eq:mrknpt2} 
and the mass-shell condition by \eqn{massnpt}. We still use the 
expansions of Eqs.~(\ref{alphb}--\ref{eq:coeffrescal}), but because of
the analytic continuation on $\kappa_1,\ldots,\kappa_{n-3}$ (which follows directly from the Eq.~(\ref{nmassnpt}) once the analytic continuation on the $s$-type invariants is established), in going
from \eqn{nmassnpt} to \eqn{massnpt}, the Lipatov vertices become complex,
\beq
\bV^{(n)}(t_{j+1},t_j,\kappa_j,\tau) = 
\left(\frac{\mu^2}{\kappa_j}\right)^{n\eps} 
V^{(n)}_{\rm phys}(t_{j+1},t_j,\kappa_j,\tau)\, ,\label{eq:posrescal}
\eeq
with
\beq
V^{(n)}_{\rm phys}(t_{j+1},t_j,\kappa_j,\tau) = e^{i\pi n\eps}\, V^{(n)}(t_{j+1},t_j,\kappa_j,\tau)\, .
\label{eq:vnlip}
\eeq

\section{The high-energy limit of  the four--, five-- and six--point MHV amplitudes}
\label{sec:456pthel}

\subsection{The four--point amplitude in multi-Regge kinematics}
\label{sec:4pthel}

For the 4--point amplitude, $g_1\,g_2\to g_3\,g_4$, the high-energy
prescription~(\ref{loopnpt}) becomes
\beq
m_4(1, 2, 3, 4) = s \left[\gs\, C(p_2,p_3,\tau) \right]
{1\over t} \left({-s\over \tau}\right)^{\alpha(t)}
\left[\gs\, C(p_1,p_4,\tau) \right]\, .\label{elasuchan}
\eeq
In order for the colour-stripped amplitude $m_4$ to be real, we take it in the
unphysical region where $s$ is negative. Then the Regge kinematics are,
\beq
-s \gg -t\, .\label{neghe}
\eeq
Using the loop expansions of the Regge trajectory (\ref{alphb}) and of the
coefficient function (\ref{fullv}), \eqn{elasuchan} can be written as \eqn{elasexpand}
for $n = 4$. Then the knowledge of the $l$-loop coefficient  $m_4^{(l)}$ allows one
to derive the $l$-loop trajectory $\alpha^{(l)}$ and coefficient function $C^{(l)}(t,\tau)$.
For example, the one-loop coefficient is given by
\beq
m_4^{(1)} = \balpha^{(1)}(t) L +\ 2 \bC^{(1)}(t,\tau)\, ,\label{4pt1l}
\eeq
with $L=\ln(-s/\tau)$, and $\balpha$ and $\bC$ rescaled as in \eqn{eq:coeffrescal}.
The one-loop trajectory is given by \eqn{alph},
\beq
\alpha^{(1)}  = \frac{2}{\eps}\, ,\label{alpha1}
\eeq
and it is the same in QCD and in MSYM. The one-loop coefficient 
function, $C^{(1)}$, has been computed in 
Ref.~\cite{Fadin:1993wh,DelDuca:1998kx,Bern:1998sc,Fadin:1992zt,Fadin:1993qb}.
In MSYM it is, to all orders in $\eps$
\bea
C^{(1)}(t,\tau) &=& 
{\psi(1+\eps) - 2\psi(-\eps) + \psi(1)\over\eps}
- \frac{1}{\eps} \ln\frac{-t}{\tau} \nn\\
&=& \frac{1}{\eps^2} \left( -2 - \eps\, \ln\frac{-t}{\tau} 
+ 3\sum_{n=1}^{\infty} \zeta_{2n}\, \eps^{2n}
+ \sum_{n=1}^{\infty} \zeta_{2n+1}\, \eps^{2n+1} \right)
\, .\label{eq:ifonel}
\eea
In fact, in the formul\ae\ that follow we shall need $C^{(1)}(t,\tau)$ through $\ord(\eps^4)$.

The two-loop coefficient of \eqn{elasexpand} with $n = 4$ is
\bea
m_4^{(2)} &=& {1\over 2} \left(\balpha^{(1)}(t)\right)^2 L^2
+ \left( \balpha^{(2)}(t) + 2\, \bC^{(1)}(t,\tau) \balpha^{(1)}(t) \right)\, L \nn\\
&+& 2\, \bC^{(2)}(t,\tau) + \left(\bC^{(1)}(t,\tau) \right)^2 \nn\\
&=& {1\over 2} \left( m_4^{(1)} \right)^2 + \balpha^{(2)}(t) L +
2\, \bC^{(2)}(t,\tau) - \left(\bC^{(1)}(t,\tau) \right)^2\, \label{exp2loopu}. 
\eea
where in the second equality we factor out the square of the
one-loop amplitude, in order to to facilitate the later comparison with
the BDS ansatz. In \eqn{exp2loopu}, $m_4^{(1)}$ must be known to $\ord(\eps^2)$.
The two-loop trajectory, $\alpha^{(2)}$, is known in full
QCD~\cite{Fadin:1995xg,Fadin:1995km,Fadin:1996tb,Blumlein:1998ib,DelDuca:2001gu}.
In MSYM, it has been computed through $\ord(\eps^0)$ 
directly~\cite{Kotikov:2000pm} and using the maximal
trascendentality principle~\cite{Kotikov:2002ab},  and through 
$\ord(\eps^2)$ directly~\cite{DelDuca:2008pj},
\beq
\alpha^{(2)} = - {2\zeta_2\over\eps} - 2\zeta_3 - 8\zeta_4\eps
+ (36\zeta_2\zeta_3 + 82\zeta_5)\eps^2 + \ord(\eps^3)\, .
\label{eq:tworegge}
\eeq
The MSYM two-loop coefficient function has been computed
through $\ord(\eps^2)$~\cite{DelDuca:2008pj},
\bea
C^{(2)}(t,\tau) &=& \frac{2}{\eps^4} + \frac{2}{\eps^3}\ln\frac{-t}{\tau}
- \left(5\zeta_2 - \frac{1}{2}\ln^2\frac{-t}{\tau}\right)\frac{1}{\eps^2} 
- \left(\zeta_3+ 2\zeta_2\ln\frac{-t}{\tau}\right)\frac{1}{\eps} 
\nn\\ &-& {55\over 4}\zeta_4 +
\left( \zeta_2\zeta_3 - 41\zeta_5 + \zeta_4\ln\frac{-t}{\tau}
\right) \eps \nn\\ &-&\left( {95\over 2}\zeta_3^2 + {1695\over 8}\zeta_6
+ (18\zeta_2\zeta_3 + 42\zeta_5) \ln\frac{-t}{\tau} \right)
\eps^2 + \ord(\eps^3) \label{eq:2loopif}\\
&=& \frac{1}{2} \left[ C^{(1)}(t,\tau) \right]^2 + \frac{\zeta_2}{\eps^2}
+ \left(\zeta_3 + \zeta_2\ln\frac{-t}{\tau}\right)\frac{1}{\eps} \nn\\
&+& \left( \zeta_3\ln\frac{-t}{\tau} - 19\zeta_4\right)
+ \left( 4\zeta_4\ln\frac{-t}{\tau} - 2\zeta_2\zeta_3 - 39\zeta_5 \right) \eps \nn\\
&-&\left( 48 \zeta_3^2 + {1773\over 8}\zeta_6
+ (18\zeta_2\zeta_3 + 41\zeta_5) \ln\frac{-t}{\tau} \right)
\eps^2 + \ord(\eps^3)\, .\nn
\eea

The three-loop coefficient is given by
\bea
m_4^{(3)} &=& 
{1\over 3!} \left(\balpha^{(1)}(t)\right)^3 L^3 
+ \balpha^{(1)}(t) \left( \balpha^{(2)}(t) + \bC^{(1)}(t,\tau)\, \balpha^{(1)}(t) \right) L^2 
\label{exp3loopu}\\
&+& \left[ \balpha^{(3)}(t) + 2\, \balpha^{(2)}(t)\, \bC^{(1)}(t,\tau)
+ \balpha^{(1)}(t) \left( 2\, \bC^{(2)}(t,\tau) + \left(\bC^{(1)}(t,\tau)\right)^2 \right) \right] L \nn\\
&+& 2\, \bC^{(3)}(t,\tau) + 2\, \bC^{(2)}(t,\tau)\, \bC^{(1)}(t,\tau) \nn\\
&=& m_4^{(2)} m_4^{(1)} - \frac{1}{3} \left(m_4^{(1)} \right)^3 \nn\\
&+& \balpha^{(3)}(t) L
+ 2\, \bC^{(3)}(t,\tau) - 2\, \bC^{(2)}(t,\tau)\, \bC^{(1)}(t,\tau)
+ \frac{2}{3} \left(\bC^{(1)}(t,\tau)\right)^3\, .\nn
\eea
In MSYM, the three-loop trajectory, $\alpha^{(3)}$, has been evaluated in
Ref.~\cite{DelDuca:2008pj,Bartels:2008ce,Drummond:2007aua,Naculich:2007ub}
through $\ord(\eps^0)$,
\beq
\alpha^{(3)} = {44\zeta_4\over 3\eps} + {40\over 3}\zeta_2\zeta_3 +
16\zeta_5 + \ord(\eps) \, .\label{eq:treregge}
\eeq
The three-loop coefficient function has been evaluated in Ref.~\cite{DelDuca:2008pj}
through $\ord(\eps^0)$ using knowledge of $m_4^{(1)}$ to $\ord(\eps^4)$, and $m_4^{(2)}$
to $\ord(\eps^2)$,
\bea
C^{(3)}(t,\tau) &=& -\frac{4}{3\eps^6} - \frac{2}{\epsilon^5} \ln\frac{-t}{\tau}
+ \left(4\zeta_2 - \ln^2\frac{-t}{\tau}\right)\frac{1}{\eps^4} \label{eq:3loopif}\\
&+& \left( 3\zeta_2 \ln\frac{-t}{\tau} - \frac{1}{6} \ln^3\frac{-t}{\tau} \right) \frac{1}{\eps^3} 
+ \left( {217\zeta_4\over 9} + \frac{\zeta_2}{2} \ln^2\frac{-t}{\tau} - \zeta_3 \ln\frac{-t}{\tau} \right)
\frac{1}{\eps^2} \nn\\
&+& \left( - {22\over 9}\zeta_2\zeta_3 + {224\over 3}\zeta_5 
-\frac{\zeta_3}{2} \ln^2\frac{-t}{\tau} +\frac{71}{12}\zeta_4 \ln\frac{-t}{\tau} \right) \frac{1}{\eps} 
\nn\\ &+&  {796\over 9}\zeta_3^2 + {211861\over 432}\zeta_6 
-\frac{5}{2} \zeta_4 \ln^2\frac{-t}{\tau} + \left(115 \zeta_5 
+ \frac{97}{3} \zeta_2 \zeta_3\right) \ln\frac{-t}{\tau} + \ord(\eps) \nn\\
&=& C^{(2)}(t,\tau)\, C^{(1)}(t,\tau) - {1\over 3} \left[C^{(1)}(t,\tau)\right]^3 \nn\\
&-& {44\over 9} \frac{\zeta_4}{\eps^2}
- \left( {40\over 9}\zeta_2\zeta_3 + {16\over 3}\zeta_5 
+ \frac{22}{3}\zeta_4 \ln\frac{-t}{\tau} \right) \frac{1}{\eps} \nn\\
&+&  \frac{3982}{27}\zeta_6  - {68\over 9}\zeta_3^2 
- \left( 8\zeta_5 + \frac{20}{3} \zeta_2 \zeta_3\right) \ln\frac{-t}{\tau} + \ord(\eps) \nn
\eea

It is straightforward to obtain the four-point amplitude in the physical region, $s \gg -t$,
by continuing Eqs.~(\ref{4pt1l}), (\ref{exp2loopu}) and (\ref{exp3loopu})
through the prescription $\ln(-s) = \ln(s) - i\pi$, for $s > 0$.
%

\subsection{The five--point amplitude in multi-Regge kinematics}
\label{sec:5pthel}

For the five-point amplitude, $g_1\,g_2\to g_3\,g_4\,g_5$, the high-energy 
prescription (\ref{loopnpt}) becomes
\beq
m_5 = s \left[g\, C(p_2,p_3,\tau) \right]\, 
{1\over t_2}\, \left({-s_2\over \tau}\right)^{\alpha(t_2)}
\left[g\,V(q_2,q_1,\kappa,\tau)\right]\, 
{1\over t_1}\, \left({-s_1\over \tau}\right)^{\alpha(t_1)}
 \left[g\, C(p_1,p_5,\tau) \right]\, , \label{loop5pt}
\eeq
%
%
where $p_4=q_2-q_1$, and
with the invariants labelled as in \sec{sec:mrkforall}, {\it i.e.}
$t_1=s_{51}$, $t_2=s_{23}$, $s_1=s_{45}$ and $s_2=s_{34}$.
In order for the amplitude $m_5$ to be real,
\eqn{loop5pt} is taken in the region where all the invariants are negative.
Thus, the multi-Regge kinematics (\ref{eq:mrkneg}) become,
\begin{equation}
-s \gg -s_{1}, -s_{2} \gg -t_1, -t_2\, .\label{eq:mrk2} 
\end{equation}
Then the mass-shell condition (\ref{nmassnpt}) for the intermediate gluon 4 is
\begin{equation}
- \kappa = {(-s_{1})\, (-s_{2})\over -s}\, ,\label{mass}
\end{equation}
where $\kappa= - |p_{4\perp}|^2$.
In the expansion of \eqn{elasexpand} for $n=5$, the knowledge of
the $l$-loop five-point amplitude in the multi-Regge kinematics (\ref{eq:mrk2}),
together with the $l$-loop trajectory $\alpha^{(l)}$ and coefficient function $C^{(l)}$,
allows one to derive the Lipatov vertex to the same accuracy.
The one-loop coefficient is
\beq
m_5^{(1)} = \balpha^{(1)}(t_1) L_1 + \balpha^{(1)}(t_2) L_2
+ \bC^{(1)}(t_1,\tau) + \bC^{(1)}(t_2,\tau) + \bV^{(1)}(t_1,t_2,\kappa,\tau)\, 
.\label{exp1loop}
\eeq
where $L_i=\ln(-s_i/\tau)$ and $i=1,2$. 
Then subtracting
the one-loop trajectory (\ref{alpha1}) and coefficient function (\ref{eq:ifonel})
from the one-loop five-point amplitude, we can derive the one-loop
Lipatov vertex. That will explicitly be done in a forthcoming publication.

In the expansion of \eqn{elasexpand} for $n=5$, the two-loop coefficient is
\bea
m_5^{(2)} &=& \frac{1}{2} \left( m_5^{(1)} \right)^2
+ \balpha^{(2)}(t_1) L_1 + \balpha^{(2)}(t_2) L_2 \label{exp2loop}\\
&+& \bC^{(2)}(t_1,\tau)  + 
\bV^{(2)}(t_1,t_2,\kappa,\tau) + \bC^{(2)}(t_2,\tau)\nn\\
&-& \frac{1}{2} \left( \bC^{(1)}(t_1,\tau) \right)^2
- \frac{1}{2} \left( \bV^{(1)}(t_1,t_2,\kappa,\tau) \right)^2
- \frac{1}{2} \left( \bC^{(1)}(t_2,\tau) \right)^2
\, ,\nn
\eea
where $m_5^{(1)}$, $\bC^{(1)}(t,\tau)$ and $ \bV^{(1)}(t_1,t_2,\kappa,\tau)$  must be known to $\ord(\eps^2)$.
Similarly, the three-loop coefficient is
\bea
m_5^{(3)} &=& 
 m_5^{(2)} m_5^{(1)} - \frac{1}{3} \left(m_5^{(1)} \right)^3  
+ \balpha^{(3)}(t_1) L_1+\balpha^{(3)}(t_2) L_2\nn \\
&+&   \bC^{(3)}(t_1,\tau) 
     +\bV^{(3)}(t_1,t_2,\kappa,\tau) 
     +\bC^{(3)}(t_2,\tau)\nn\\
&-& \bC^{(2)}(t_1,\tau)\, \bC^{(1)}(t_1,\tau)
  - \bV^{(2)}(t_1,t_2,\kappa,\tau) \bV^{(1)}(t_1,t_2,\kappa,\tau)
   -\bC^{(2)}(t_2,\tau)\, \bC^{(1)}(t_2,\tau) \nn\\
&+& \frac{1}{3} \left(\bC^{(1)}(t_1,\tau)\right)^3
   +\frac{1}{3} \left(\bV^{(1)}(t_1,t_2,\kappa,\tau) \right)^3
   +\frac{1}{3} \left(\bC^{(1)}(t_2,\tau)\right)^3\, .\label{m53ite}
\eea
Here, to find $m_5^{(3)}$ to $\ord(\eps^0)$, $m_5^{(1)}$, $\bC^{(1)}(t,\tau)$ and $ \bV^{(1)}(t_1,t_2,\kappa,\tau)$  must be
known to $\ord(\eps^4)$ while 
$m_5^{(2)}$, $\bC^{(2)}(t,\tau)$ and $ \bV^{(2)}(t_1,t_2,\kappa,\tau)$  
must be known to $\ord(\eps^2)$.

%

It is straightforward to obtain the amplitudes in the physical region where
$s, s_1, s_2$ are positive and $t_1, t_2$ are negative, and where
the multi-Regge kinematics are
\begin{equation}
s \gg s_{1},\ s_{2} \gg -t_1,\ -t_2\, .\label{eq:posmrk5pt} 
\end{equation}
and the mass-shell condition is
\begin{equation}
\kappa = {s_{1}\, s_{2}\over s}\, ,\label{mass5pt}
\end{equation}
by continuing \eqns{exp1loop}{exp2loop} 
through the prescriptions $\ln(-s_j) = \ln(s_j) - i\pi$, for $s_j > 0$ and $j=1, 2$
and $\ln(-\kappa) = \ln(\kappa) - i\pi$, for $\kappa > 0$, which implies
\eqn{eq:posrescal} for the Lipatov vertex.
%
%

\subsection{The six-point amplitude in multi-Regge kinematics}
\label{sec:6pthel}

For the six-gluon amplitude, $g_1\,g_2\to g_3\,g_4\,g_5\,g_6$,
the high-energy prescription (\ref{loopnpt}) in the Euclidean region becomes
\bea
\lefteqn{ m_6 = s \left[g\, C(p_2,p_3,\tau) \right]\, 
{1\over t_3}\, \left({-s_3\over \tau}\right)^{\alpha(t_3)}\, 
\left[g\, V(q_2,q_3,\kappa_2,\tau)\right] } \nn\\
&\times& {1\over t_2}\, \left({-s_2\over \tau}\right)^{\alpha(t_2)}\, 
\left[g\, V(q_1,q_2,\kappa_1,\tau)\right]\,
{1\over t_1}\, \left({-s_1\over \tau}\right)^{\alpha(t_1)}
\left[g\, C(p_1,p_6,\tau) \right]\, .\label{eq:hepresc6pt}
\eea
with $t_1=s_{61}$, $t_2=s_{234}$ and $t_3=s_{23}$, $s_1=s_{56}$, $s_2=s_{45}$ 
and $s_3=s_{34}$. In order for $m_6$ to be real, we take
\eqn{eq:hepresc6pt} in the unphysical region where the invariants 
$s, s_1, s_2, s_3, t_1, t_2, t_3$ are all negative,
where the multi-Regge kinematics are,
\begin{equation}
-s \gg -s_{1}, -s_{2}, -s_{3} \gg -t_1, -t_2, -t_3\, ,\label{eq:mrkneg6pt} 
\end{equation}
and the on-shell conditions (\ref{nmassnpt}) are,
\begin{equation}
- \kappa_1 = {(-s_1)\, (-s_2)\over (-s_{456}) }\, , \qquad 
- \kappa_2 = {(-s_2)\, (-s_3)\over (-s_{345}) } \, 
,\label{negmass6pt}
\end{equation}
with $\kappa_1= - |p_{5\perp}|^2$ and $\kappa_2= - |p_{4\perp}|^2$.
Because in \eqn{eq:hepresc6pt} no new vertex or coefficient function occurs
with respect to \eqn{loop5pt},
in the expansion of \eqn{elasexpand} for $n=6$, the knowledge of
the $l$-loop trajectory $\alpha^{(l)}$, the coefficient function $C^{(l)}$,
and the Lipatov vertex $V^{(l)}$ allow one to derive
the $l$-loop six-point amplitude in the multi-Regge kinematics. The one-loop coefficient is
\bea
m_6^{(1)} &=& \balpha^{(1)}(t_1) L_1 + \balpha^{(1)}(t_2) L_2
+ \balpha^{(1)}(t_3) L_3 \nn\\
&+& \bC^{(1)}(t_1,\tau) + \bC^{(1)}(t_3,\tau) 
+ \bV^{(1)}(t_1,t_2,\kappa_1,\tau) + \bV^{(1)}(t_2,t_3,\kappa_2,\tau) 
.\label{exp1loop6pt}
\eea
with $L_i=\ln(-s_i/\tau)$ and $i=1,2,3$. The two-loop coefficient is
\bea
m_6^{(2)} &=& \frac{1}{2} \left( m_6^{(1)} \right)^2
+ \balpha^{(2)}(t_1) L_1 + \balpha^{(2)}(t_2) L_2 + \balpha^{(2)}(t_3) L_3 
\label{exp2loop6pt}\\
&+& \bC^{(2)}(t_1,\tau) + \bC^{(2)}(t_3,\tau) + 
\bV^{(2)}(t_1,t_2,\kappa_1,\tau) 
+ \bV^{(2)}(t_2,t_3,\kappa_2,\tau) \nn\\
&-& \frac{1}{2} \left( \bC^{(1)}(t_1,\tau) \right)^2
- \frac{1}{2} \left( \bC^{(1)}(t_3,\tau) \right)^2\nn \\
&-& \frac{1}{2} \left( \bV^{(1)}(t_1,t_2,\kappa_1,\tau) \right)^2
- \frac{1}{2} \left( \bV^{(1)}(t_2,t_3,\kappa_2,\tau) \right)^2\, ,\nn
\eea
where $m_6^{(1)}$,  $\bC^{(1)}(t,\tau)$ and $ \bV^{(1)}(t_1,t_2,\kappa,\tau)$ must be known to $\ord(\eps^2)$.
Similarly, the three-loop coefficient is
\bea
m_6^{(3)} &=& 
 m_6^{(2)} m_6^{(1)} - \frac{1}{3} \left(m_6^{(1)} \right)^3  
+ \balpha^{(3)}(t_1) L_1+\balpha^{(3)}(t_3) L_2\nn \\
&+&  \bC^{(3)}(t_1,\tau) 
    +\bV^{(3)}(t_1,t_2,\kappa_1,\tau) 
    +\bV^{(3)}(t_2,t_3,\kappa_2,\tau) 
    +\bC^{(3)}(t_3,\tau)\nn\\
&-& \bC^{(2)}(t_1,\tau)\, \bC^{(1)}(t_1,\tau)
  - \bV^{(2)}(t_1,t_2,\kappa_1,\tau) \bV^{(1)}(t_1,t_2,\kappa_1,\tau)\nn\\
&-& \bV^{(2)}(t_2,t_3,\kappa_2,\tau) \bV^{(1)}(t_2,t_3,\kappa_2,\tau)
   -\bC^{(2)}(t_3,\tau)\, \bC^{(1)}(t_3,\tau) \nn\\
&+& \frac{1}{3} \left(\bC^{(1)}(t_1,\tau)\right)^3
+\frac{1}{3} \left(\bC^{(1)}(t_3,\tau)\right)^3\nn\\
   &+&\frac{1}{3} \left(\bV^{(1)}(t_1,t_2,\kappa_1,\tau) \right)^3
   +\frac{1}{3} \left(\bV^{(1)}(t_2,t_3,\kappa_2,\tau) \right)^3
   \, .\label{eq:m63exp}
\eea
Here, $m_6^{(1)}$, $\bC^{(1)}(t,\tau)$ and $ \bV^{(1)}(t_1,t_2,\kappa,\tau)$  are needed
to $\ord(\eps^4)$ while 
$m_6^{(2)}$, $\bC^{(2)}(t,\tau)$ and $ \bV^{(2)}(t_1,t_2,\kappa,\tau)$  
must be known to $\ord(\eps^2)$.

It is straightforward to obtain the amplitudes in the physical region where
$s, s_1, s_2, s_3$ are positive and $t_1, t_2, t_3$ are negative, where
the multi-Regge kinematics are
\begin{equation}
s \gg s_1, s_2, s_3 \gg -t_1, -t_2, -t_3\, ,\label{eq:mrk6pt} 
\end{equation}
and the mass-shell conditions for gluons 4 and 5, emitted along
the $t$ channel, are
\begin{equation}
\kappa_1 = {s_{1}\, s_{2}\over s_{456}} \, ,\qquad \kappa_2 = {s_{2}\, s_{3}\over s_{345}} \, 
,\label{mass6pt}
\end{equation}
by continuing \eqns{exp1loop6pt}{exp2loop6pt} 
through the prescriptions $\ln(-s_j) = \ln(s_j) - i\pi$, for $s_j > 0$ with $j=1, 2,3$,
and $\ln(-\kappa_i) = \ln(\kappa_i) - i\pi$, for $\kappa_i > 0$, with $i=1,2$.

\section{The Bern-Dixon-Smirnov ansatz in multi-Regge kinematics}
\label{sec:bdsmrk}

The BDS ansatz prescribes that the $n$-gluon MHV amplitude be written as,
\bea
m_n &=& m_n^{(0)} \left[ 1 + \sum_{L=1}^\infty a^L M_n^{(L)}(\eps) \right] 
\nn\\ &=& m_n^{(0)} 
\exp\left[ \sum_{l=1}^\infty a^l \left( f^{(l)}(\eps) 
M_n^{(1)}(l\eps) + Const^{(l)} + E_n^{(l)}(\eps)\right)\right]\, 
,\label{eq:bds1}
\eea
where
\beq
a = {2\gs^2 N\over (4\pi)^{2-\eps}} e^{-\gamma\eps}
\eeq
is the 't-Hooft gauge coupling, and with
\beq
f^{(l)}(\eps) = f^{(l)}_0 + \eps f^{(l)}_1 + \eps^2 f^{(l)}_2\, 
,\label{eq:flfunct}
\eeq
where $f^{(1)}(\eps)=1$, and
$f^{(l)}_0$ is proportional to the $l$-loop cusp anomalous 
dimension~\cite{Korchemsky:1987wg}, 
$\hat{\gamma}_K^{(l)} = 4f^{(l)}_0$, which has been conjectured to all orders of 
$a$~\cite{Beisert:2006ez} and computed to 
$\ord(a^4)$~\cite{Bern:2006ew,Cachazo:2006az}, and $f^{(l)}_1$
is related to the soft anomalous 
dimension~\cite{Magnea:1990zb,Sterman:2002qn}, 
${\cal G}_0^{(l)} = 2f^{(l)}_1/l$, and is known to
$\ord(a^3)$~\cite{Bern:2005iz}. In \eqn{eq:bds1},
$Const^{(l)}$ are constants, and $E_n^{(l)}(\eps)$ are $\ord(\eps)$ 
contributions, with $Const^{(1)}=0$ and $E_n^{(1)}(\eps)=0$,
and $M_n^{(L)}(\eps)$ is the $L$-loop colour-stripped
amplitude rescaled by the tree amplitude. In the convention and notation 
of \eqn{elasexpand}, the rescaled coupling (\ref{rescal}) is related to $a$ by,
\beq
a = 2 G(\eps)\tgs^2 
\eeq
with
\beq
G(\eps) = {e^{-\gamma\eps}\ \Gamma(1-2\epsilon)\over
\Gamma(1+\epsilon)\, \Gamma^2(1-\epsilon)} = 1 + \ord(\eps^2)\, .
\eeq
Thus, the $n$-gluon amplitude is given by,
\beq
a^L M_n^{(L)}(\eps) = \left( \frac{a}{2G(\eps)}\right)^L
m_n^{(L)}(\eps)\, ,\label{eq:ourm}
\eeq
and the BDS ansatz (\ref{eq:bds1}) becomes
\bea
m_n &=& m_n^{(0)} \left[ 1 + \sum_{L=1}^\infty {\bar\gs}^{2L}(t) 
m_n^{(L)}(\eps) \right] \nn\\ &=& m_n^{(0)} 
\exp\left[ \sum_{l=1}^\infty {\bar\gs}^{2l}(t) \left( 2G(\eps)\right)^l 
\left( f^{(l)}(\eps) {m_n^{(1)}(l\eps)\over 2G(l\eps)}
+ Const^{(l)} +E_n^{(l)}(\eps) \right)\right]\, .\label{eq:bdsddg}
\eea

\subsection{Amplitudes with four or five gluons}

Substituting the one-loop four-point amplitude (\ref{4pt1l}) in
\eqn{eq:bdsddg} and comparing
with the expansion (\ref{elasexpand}) for $n=4$ of the high-energy 
prescription (\ref{elasuchan}), we determine the Regge trajectory from
the coefficient of the single logarithm~\cite{DelDuca:2008pj},
\bea
\alpha^{(2)}(\eps) &=& 2\, f^{(2)}(\eps)\, \alpha^{(1)}(2\eps) + \ord(\eps)\, , \label{eq:alphabds}\\
\alpha^{(3)}(\eps) &=& 4\, f^{(3)}(\eps)\, \alpha^{(1)}(3\eps) + \ord(\eps)\, ,\nn
\eea
with $\alpha^{(1)}$ given in \eqn{alpha1}, and in general
\beq
\alpha^{(l)}(\eps) = 2^{l-1}\, f^{(l)}(\eps)\, \alpha^{(1)}(l\eps) + \ord(\eps)\, .\label{eq:alphagen}
\eeq
From \eqn{eq:alphagen}, we see that to $\ord(\eps^0)$
only the first two terms of the $f^{(l)}(\eps)$ function (\ref{eq:flfunct})
enter the evaluation of the Regge trajectory.
Using the $f^{(2)}$ and $f^{(3)}$ functions~\cite{Bern:2005iz},
\bea
f^{(2)}(\eps) &=& - \zeta_2 - \zeta_3\eps - \zeta_4\eps^2\, ,\nn\\
f^{(3)}(\eps) &=& {11\over 2} \zeta_4 + (6\zeta_5 + 5\zeta_2\zeta_3)\eps
+ (c_1\zeta_6 + c_2\zeta_3^2)\eps^2\, ,\label{eq:ffunct} 
\eea
we see that \eqn{eq:alphabds} agrees with \eqns{eq:tworegge}{eq:treregge}
to $\ord(\eps^0)$. The constants $c_1, c_2$ are known only 
numerically~\cite{Spradlin:2008uu}, but they
do not enter the evaluation of the Regge trajectory.

\eqn{eq:bdsddg} implies the iterative structure of the two-loop $n$-gluon
amplitude given in \eqn{eq:ite2bds}, which we report here in our 
convention~(\ref{eq:ourm}) for the coupling,
\beq
m_n^{(2)}(\eps) = {1\over 2} \left[m_n^{(1)}(\eps)\right]^2
+ {2\,G^2(\eps)\over G(2\eps)} f^{(2)}(\eps)\, m_n^{(1)}(2\eps) 
+ 4\, Const^{(2)} + \ord(\eps)\, ,\label{eq:ite2}
\eeq
with $Const^{(2)}= -\zeta_2^2/2$, and where the one-loop 
amplitude, $m_n^{(1)}(\eps)$, must be
known to $\ord(\eps^2)$. \eqn{eq:ite2} has been shown to be correct for
$n= 4$~\cite{Anastasiou:2003kj} and $n=5$~\cite{Bern:2006vw,Cachazo:2008vp} for
general kinematics.

Using the iterative structure (\ref{eq:ite2}) for the four-point amplitude,
it is possible to express the two-loop coefficient function in terms of
the one-loop coefficient function. In fact, comparing 
\eqn{eq:ite2} with $n= 4$ to the two-loop factorization of the
four-point amplitude in the multi-Regge kinematics~(\ref{exp2loopu}), 
we find the following iterative structure
\beq
C^{(2)}(t,\tau,\eps) = {1\over 2} \left[ C^{(1)}(t,\tau,\eps)\right]^2
+ {2\,G^2(\eps)\over G(2\eps)} f^{(2)}(\eps)\, C^{(1)}(t,\tau,2\eps) 
+ 2\, Const^{(2)}
+ \ord(\eps)\, ,\label{eq:ifite2}
\eeq
where, to compute the two-loop coefficient function $C^{(2)}(t,\tau,\eps)$ to $\ord(\eps^0)$, the one-loop coefficient function, $C^{(1)}(t,\tau,\eps)$, 
is needed to $\ord(\eps^2)$. \eqn{eq:ifite2} agrees with 
\eqn{eq:2loopif} to $\ord(\eps^0)$.

Similarly,  the iterative structure (\ref{eq:ite2}) for the five-point amplitude,
means we can also 
express the two-loop Lipatov vertex in terms of the one-loop Lipatov vertex.
Comparing \eqn{eq:ite2} with $n= 5$ to the two-loop factorization of the
five-point amplitude~(\ref{exp2loop}), and using \eqns{eq:alphabds}{eq:ifite2},
we obtain
\beq
V^{(2)}(t_1,t_2,\kappa,\tau,\eps) = {1\over 2} \left[ V^{(1)}(t_1,t_2,\kappa,\tau,\eps)\right]^2
+ {2\,G^2(\eps)\over G(2\eps)} f^{(2)}(\eps)\, V^{(1)}(t_1,t_2,\kappa,\tau,2\eps) 
+ \ord(\eps)\, ,\label{eq:2llipver}
\eeq
where, to compute $V^{(2)}(t_1,t_2,\kappa,\tau,\eps)$ to $\ord(\eps^0)$, $V^{(1)}(t_1,t_2,\kappa,\tau,\eps)$, must be known 
through $\ord(\eps^2)$.
Of course, \eqn{eq:ite2} with $n= 5$ requires the knowledge of
the one-loop five-point amplitude, $m_5^{(1)}(\eps)$, through 
$\ord(\eps^2)$\footnote{We shall provide the details of $m_5^{(1)}(\eps)$ to that accuracy,
in fact to all orders in $\eps$, in a forthcoming publication~\cite{us}.}, but once $V^{(1)}$ is known 
through $\ord(\eps^2)$, the two-loop Lipatov vertex can be determined by \eqn{eq:2llipver}
without knowing explicitly the two-loop five-point amplitude. In fact, once evaluated,
$V^{(2)}$ can be used, together with $C^{(2)}$ and $\alpha^{(2)}$, in \eqn{exp2loop}
to determine the two-loop five-point amplitude in the multi-Regge kinematics.

The iterative structure of the three-loop $n$-gluon amplitude is,
\beq
m_n^{(3)}(\eps) = m_n^{(2)}(\eps)\, m_n^{(1)}(\eps)
- {1\over 3} \left[m_n^{(1)}(\eps)\right]^3
+ {4\,G^3(\eps)\over G(3\eps)} f^{(3)}(\eps)\, m_n^{(1)}(3\eps) 
+ 8\, Const^{(3)} + \ord(\eps)\, ,\label{eq:ite3}
\eeq
where $m_n^{(1)}(\eps)$ and $m_n^{(2)}(\eps)$ must be known to
$\ord(\eps^4)$ and $\ord(\eps^2)$, respectively, and with
\beq
Const^{(3)} = \left( {341\over 216} + {2\over 9} c_1\right) \zeta_6
+ \left( -{17\over 9} + {2\over 9} c_2\right) \zeta_3^2\, .\label{eq:cost3}
\eeq
\eqn{eq:ite3} has been shown to be correct for $n=4$~\cite{Bern:2005iz}.

Comparing \eqn{eq:ite3} with $n= 4$ to the three-loop factorisation of the
four-point amplitude in the multi-Regge kinematics~(\ref{exp3loopu}),
we obtain the three-loop iteration of the coefficient function,
\bea
C^{(3)}(t,\tau,\eps) &=& C^{(2)}(t,\tau,\eps)\, C^{(1)}(t,\tau,\eps)
 - {1\over 3} \left[C^{(1)}(t,\tau,\eps)\right]^3 \nn\\
&+& {4\,G^3(\eps)\over G(3\eps)} f^{(3)}(\eps)\, C^{(1)}(t,\tau,3\eps)
+ 4\, Const^{(3)} + \ord(\eps)\, .\label{eq:ifite3}
\eea
The constants $c_1, c_2$ cancel when \eqns{eq:ffunct}{eq:cost3} are used
in \eqns{eq:ite3}{eq:ifite3}.
Using the two-loop coefficient function to $\ord(\eps^2)$ (\ref{eq:2loopif}),
and the one-loop coefficient function to $\ord(\eps^4)$ (\ref{eq:ifonel}),
we see that \eqn{eq:ifite3} is in agreement with \eqn{eq:3loopif} to 
$\ord(\eps^0)$.

Comparing \eqn{eq:ite3} with $n= 5$ to the three-loop factorisation of the
five-point amplitude (\ref{m53ite}), we obtain the three-loop iteration of the
Lipatov vertex,
\bea
V^{(3)}(t_1,t_2,\kappa,\tau,\eps) &=& 
V^{(2)}(t_1,t_2,\kappa,\tau,\eps) V^{(1)}(t_1,t_2,\kappa,\tau,\eps)
- {1\over 3} \left[ V^{(1)}(t_1,t_2,\kappa,\tau,\eps)\right]^3 \nn\\
&+& {4\,G^3(\eps)\over G(3\eps)} f^{(3)}(\eps)\, V^{(1)}(t_1,t_2,\kappa,\tau,3\eps) 
+ \ord(\eps)\, .\label{eq:3llipver}
\eea

\subsection{Amplitudes with six or more gluons}

In the two-loop expansion of the six-point amplitude~(\ref{exp2loop6pt}), 
no new vertices or coefficient functions occur. Thus, using the explicit expressions
of $V^{(2)}$, $C^{(2)}$ and $\alpha^{(2)}$ in \eqn{exp2loop6pt}, 
one can assemble the two-loop six-point amplitude in the multi-Regge kinematics.
However, even without knowing the explicit expression of the two-loop Lipatov 
vertex~(\ref{eq:2llipver}), it is easy to see by substitution that 
the iterative structure of Eqs.~(\ref{eq:alphabds}),
(\ref{eq:ifite2}) and (\ref{eq:2llipver}) ensures that the six-point amplitude~(\ref{exp2loop6pt})
fulfils the two-loop iterative formula (\ref{eq:ite2}) for $n=6$. Furthermore, the expression
has the correct analytic properties in the physical region where
$s, s_1, s_2, s_3$ are positive and $t_1, t_2, t_3$ are negative. 

Because no new vertices or coefficient functions occur in the two-loop expansion
of \eqn{loopnpt} even for $n=7$ or higher, we conclude that the two-loop expansion
of \eqn{loopnpt} fulfils the two-loop iterative formula (\ref{eq:ite2}), and thus the BDS
ansatz, for any $n$. Thus, the multi-Regge kinematics
are not able to resolve the BDS-ansatz discrepancy, {\it i.e.} the quantity $R_n^{(2)}$
(\ref{eq:discr}) vanishes in the multi-Regge kinematics, for any $n$.

The same arguments can be repeated for three-loop case: in the three-loop expansion of
the six-point amplitude (\ref{eq:m63exp}) no new vertices or coefficient functions occur.
Thus, using the explicit expressions
of $V^{(3)}$, $C^{(3)}$ and $\alpha^{(3)}$ in \eqn{eq:m63exp}
one can assemble the three-loop six-point amplitude in the multi-Regge kinematics.
However, even without knowing the explicit expression of the three-loop Lipatov 
vertex~(\ref{eq:3llipver}), it is easy to see by substitution that 
the iterative structure of Eqs.~(\ref{eq:alphabds}),
(\ref{eq:ifite3}) and (\ref{eq:3llipver}) ensures that the six-point amplitude~(\ref{eq:m63exp})
fulfils the three-loop iterative formula (\ref{eq:ite3}) for $n=6$.
Because no new vertices or coefficient functions occur in the three-loop expansion
of \eqn{loopnpt} for $n=7$ or higher, the three-loop expansion
of \eqn{loopnpt} fulfils the three-loop iterative formula (\ref{eq:ite3}), and thus the BDS
ansatz, for any $n$. Thus, also the quantity $R_n^{(3)}$
(\ref{eq:discr}) vanishes in the multi-Regge kinematics, for any $n$.
Clearly, the same thing is to occur with the iterative structure of the $l$-loop
$n$-gluon amplitude for $l\ge 4$. We conclude that $R_n^{(l)}$
vanishes in the multi-Regge kinematics for any $l$ and $n$.
The $l$-loop $n$-gluon amplitudes in the multi-Regge kinematics are in complete
agreement with the BDS ansatz, therefore they are not able to resolve the violations of
the ansatz for $n\ge 6$.

In Ref.~\cite{Drummond:2007au, Bern:2008ap} it was argued that the remainder function~(\ref{eq:discr}) for $n=6$ is a function of the three conformal cross-ratios
\beq
u_1 = \frac{s_{12}\, s_{45}}{s_{345}\, s_{456}}\, ,\qquad
u_2 = \frac{s_{23}\, s_{56}}{s_{234}\, s_{456}}\, ,\qquad
u_3 = \frac{s_{34}\, s_{61}}{s_{234}\, s_{345}}\, .\label{thrinvar}
\eeq
Using the notation of \sec{sec:mrkforall} and the
results of \sec{sec:6ptmrkinvar}, we note that
in the multi-Regge kinematics (\ref{eq:mrk6pt})
the conformal invariants (\ref{thrinvar}) become~\cite{Brower:2008nm, Brower:2008ia}
\beq
u_1  \simeq 1\, ,\qquad 
u_2 = \frac{t_3 \kappa_1}{t_2 s_2} \simeq \ord\left(\frac{t}{s}\right)\, ,\qquad
u_3 = \frac{t_1 \kappa_2}{t_2 s_2} \simeq \ord\left(\frac{t}{s}\right)\, ,\label{thrinvarmrk}
\eeq
thus $u_1$ is close to 1, while $u_2$ and $u_3$ are very small and are in fact sub-leading
in the multi-Regge kinematics.

\section{Proof of BDS ansatz in multi-Regge kinematics}
\label{sec:proof}

In the previous section, we derived iterative relations for the three building blocks that occur in the
multi-Regge factorisation of gluonic amplitudes, the Regge trajectory, the coefficient functions and the
Lipatov vertex. We argued that the high-energy prescription implied  that the six-gluon amplitude also
satisfies the BDS ansatz (in the restricted kinematics where the high energy prescription is valid).  In
this section, we are going to prove that in in the Euclidean region the BDS ansatz is fully consistent 
with multi-Regge factorisation (the proof for the physical region is similar).
In particular, we show that, if BDS holds true for four- and five-point amplitudes, then
it also holds true for any $n$-gluon amplitude (in multi-Regge kinematics). 

We start by deriving exponentiated forms for the coefficient functions and the
Lipatov vertex. 
If the BDS ansatz holds true for the four-point amplitude, then we can immediately
insert the tree- and one-loop four-gluon
amplitudes in multi-Regge kinematics  
\beq\begin{split}\label{eq:m41}
m_4^{(0)}=&g^2 C^{(0)}(p_2,p_3)\,\frac{s}{t}\,C^{(0)}(p_1,p_4),\\
m_4^{(1)}(l\eps)=&2\bC^{(1)}(t,\tau,l\eps) + \bar\alpha^{(1)}(t,l\eps)\ln\left(\frac{-s}{\tau}\right),
\end{split}
\eeq
into \eqn{eq:bdsddg}, such that
\bea\label{eq:m4exp}
m_4
&=&\,g^2 C^{(0)}(p_2,p_3)\,\frac{s}{t}\,C^{(0)}(p_1,p_4) \left(\frac{-s}{\tau}\right)^{\sum_{l=1}^{\infty}\,\tgs^{2l}\,
2^{l-1}\frac{G^l(\eps)}{G(l\eps)}\,f^{(l)}(\eps)\,\bar\alpha^{(1)}(t,l\eps)}\nonumber \\
&&\,\times\exp\,2\sum_{l=1}^{\infty}\,\tgs^{2l}\, 2^{l-1}G^l(\eps)\left(\frac{f^{(l)}(\eps)}{G(l\eps)}\,\bC^{(1)}(t,\tau,l\eps)+Const^{(l)}+E_4^{(l)}(\eps)\right).
\eea
Comparing \eqn{eq:m4exp} to the general form of the high energy prescription of \eqn{elasuchan}, 
we can easily identify the all-orders forms of the Regge trajectory
\beq\label{eq:aExp}
\alpha(t,\eps)=\sum_{l=1}^{\infty}\,\tgs^{2l}\, 2^{l-1}\frac{G^l(\eps)}{G(l\eps)}\,f^{(l)}(\eps)\,\bar\alpha^{(1)}(t,l\eps),
\eeq
and the coefficient function,
\beq\begin{split}\label{eq:CExp}
C(p_i,p_j,&\tau,\eps)=\\ 
&\,C^{(0)}(p_i,p_j)
\,\exp\,\sum_{l=1}^{\infty}\,\tgs^{2l}\, 2^{l-1}
G^l(\eps)\left(\frac{f^{(l)}(\eps)}{G(l\eps)}\,\bC^{(1)}(t,\tau,l\eps)+Const^{(l)}+E_4^{(l)}(\eps)\right),
\end{split}
\eeq
where in the last equation $t=(p_i+p_j)^2$. Note that expanding \eqn{eq:aExp} and \eqn{eq:CExp} in the rescaled couplings reproduces the 
explicit forms for the two-, and three-loop
iterative expressions given in \eqn{eq:alphabds} and \eqns{eq:ifite2}{eq:ifite3} respectively. Furthermore, \eqn{eq:aExp} is in agreement up to $\ord(\eps)$ with \eqn{eq:alphagen}, which expresses the $l$-loop Regge trajectory in terms of the function $f^{(l)}$ appearing in the BDS ansatz.

We can now repeat the argument for $m_5$ and, by reusing \eqn{eq:aExp} and \eqn{eq:CExp},  extract the corresponding formula for the Lipatov vertex,
\bea\label{eq:m5exp}
m_5&=&\,g^2\, C(p_2,p_3,\tau,\eps)\,\frac{s}{t_1\,t_2}\,V^{(0)}(q_2,q_1)\,C(p_1,p_5,\tau,\eps)\nonumber \\
&&\,\times
\left(\frac{-s_1}{\tau}\right)^{\alpha(t_1,\eps)}\,
\left(\frac{-s_2}{\tau}\right)^{\alpha(t_2,\eps)}\nonumber\\
&&\,\times\exp\,\sum_{l=1}^{\infty}\,\tgs^{2l}\, 2^{l}G^l(\eps)\left(\frac{f^{(l)}(\eps)}{2G(l\eps)}
\,\bV^{(1)}(t_2,t_1,\kappa_1,\tau,l\eps) + 
E_5^{(l)}(\eps) - E_4^{(l)}(\eps)\right). 
\eea
Comparing with \eqn{loop5pt}, we find
\bea
\label{eq:VExp}
V(q_2,q_1,\kappa,\eps)&=&V^{(0)}(q_2,q_1)\,\nonumber\\
&\times&\exp\,\sum_{l=1}^{\infty}\,\tgs^{2l}\,
2^{l}G^l(\eps)\left(\frac{f^{(l)}(\eps)}{2G(l\eps)}\,\bV^{(1)}(t_2,t_1,\kappa_1,\tau,l\eps) + E_5^{(l)}(\eps) -
E_4^{(l)}(\eps)\right).\nonumber \\
\eea
As before,  expanding \eqn{eq:VExp} in the rescaled couplings reproduces the 
explicit forms for the two-, and three-loop
iterative expressions given in \eqns{eq:2llipver}{eq:3llipver}.

We now turn to the generic case. Consider an $n$-gluon amplitude in multi-Regge kinematics which satisfies \eqn{loopnqmr}.
Inserting the exponentiated expressions for the Regge trajectory \eqn{eq:aExp}, the coefficient functions \eqn{eq:CExp} and the Lipatov verticx \eqn{eq:VExp}, 
we find
\bea
m_n&=&m_n^{(0)}\,\exp\,\sum_{l=1}^{\infty}\,\tgs^{2l}\, 2^{l}G^l(\eps)\Bigg[\frac{f^{(l)}(\eps)}{2G(l\eps)}\,\Bigg(\bC^{(1)}(t_1,\tau,l\eps)+\bC^{(1)}(t_{n-3},\tau,l\eps) \nonumber \\
&&\qquad + \sum_{k=1}^{n-3}\bar\alpha^{(1)}(t_k,l\eps)\ln\left(\frac{-s_k}{\tau}\right)\nonumber \\
 && \qquad+\sum_{k=1}^{n-4}\bV^{(1)}(t_{k+1},t_k,\kappa_k,\tau,l\eps)\Bigg)\nonumber \\
 &&\qquad+Const^{(l)}+E_4^{(l)}(\eps) + (n-4)\big(E_5^{(l)}(\eps)-E_4^{(l)}(\eps)\big) \Bigg].
 \eea
The expression inside the brackets can now be easily identified as the one-loop amplitude in multi-Regge kinematics,
\bea
m_n^{(1)}(l\eps)&=&\,\bC^{(1)}(t_1,\tau,l\eps)+\bC^{(1)}(t_{n-3},\tau,l\eps) 
+ \sum_{k=1}^{n-3}\bar\alpha^{(1)}(t_k,l\eps)\ln\left(\frac{-s_k}{\tau}\right)\nonumber\\
&&\, +\sum_{k=1}^{n-4}\bV^{(1)}(t_{k+1},t_k,\kappa_k,\tau,l\eps),
 \eea
 and so we recover
 \beq
m_n=m_n^{(0)}\,\exp\,\sum_{l=1}^{\infty}\,\tgs^{2l}\, 2^{l}G^l(\eps)\Bigg(\frac{f^{(l)}(\eps)}{2G(l\eps)}\,m_n^{(1)}(l\eps)
+Const^{(l)}+\ord(\eps)\Bigg),
 \eeq
 \emph{i.e.} $m_n$ satisfies the BDS ansatz up to $\ord(\eps)$.

\section{Quasi-multi-Regge kinematics}
\label{sec:quasi}

\subsection{Amplitudes in the quasi-multi-Regge kinematics with a pair at either end of the ladder}
\label{sec:ampnqmr}

\begin{figure}[!t]
 \begin{fmffile}{quasimr}
               (a)  \parbox{70mm}{ \begin{fmfgraph*}(150,250)
                  \fmfstraight
                     \fmfleft{p1,pdn1,pdn2,pd5,pd4,p2}
                     \fmfright{x1,x2,x3,x5,x6,x7}
                     \fmf{phantom}{p1,u1,v1n,u2,pn,u3,x1}
                     \fmf{phantom}{p2,o1,v23,o2,p3,o3,x7}
                     \fmffreeze
                     \fmf{phantom}{pn,pn1,pn2,p5,p4,p3}
                     \fmffreeze
                     \fmf{gluon,label=$p_2$,label.side=left,l.d=0.03w}{p2,v23}
                     \fmf{gluon,label=$p_3$,label.side=left,l.d=0.03w}{v23,p3}
                     \fmf{phantom}{v23,v4,v5,vn2,vn1,v1n}
                     \fmffreeze
                     \fmfv{decor.shape=circle,decor.filled=shaded,decor.size=0.19w,fore=green}{v23}
                     \fmfv{decor.shape=circle,decor.filled=shaded,decor.size=.09w,fore=green}{v4}
                     \fmfv{decor.shape=circle,decor.filled=shaded,decor.size=.09w,fore=green}{v5}
                     \fmfv{decor.shape=circle,decor.filled=shaded,decor.size=.09w,fore=green}{vn2}
                     \fmfv{decor.shape=circle,decor.filled=shaded,decor.size=.09w,fore=green}{vn1}
                     \fmfv{decor.shape=circle,decor.filled=shaded,decor.size=.09w,fore=green}{v1n}
                     \fmf{zigzag,label=$q_{n-4}$,label.side=right,l.d=0.055w}{v23,v4}
                     \fmf{zigzag,label=$q_{n-5}$,label.side=right,l.d=0.055w}{v4,v5}
                     \fmf{zigzag,label=$q_{2}$,label.side=right,l.d=0.055w}{vn2,vn1}
                     \fmf{zigzag,label=$q_{1}$,label.side=right,l.d=0.055w}{vn1,v1n}
                     \fmf{gluon,label=$p_1$,label.side=right,l.d=0.055w}{p1,v1n}
                     \fmf{gluon,label=$p_{n}$,label.side=right,l.d=0.055w}{v1n,pn}
                     \fmffreeze
                     \fmf{gluon,label=$p_5$,label.side=left,l.d=0.03w}{v4,p4}
                     \fmf{gluon,label=$p_6$,label.side=left,l.d=0.03w}{v5,p5}
                     \fmf{gluon,label=$p_{n-2}$,label.side=left,l.d=0.03w}{vn2,pn2}
                     \fmf{gluon,label=$p_{n-1}$,label.side=left,l.d=0.03w}{vn1,pn1}
                     \fmffreeze
                     \fmf{phantom}{u3,ou4,ou5,oun2,oun1,o3}
                     \fmffreeze
                     \fmf{phantom}{o3,ox1,ox2,ox3,ox4,ox5,ox6,ox7,ox8,ox9,ox10,oun1}
                     \fmffreeze
                     \fmf{plain,tension=0.2,left=0.3,label=$s_{n-4}$}{ox4,ox10}
                     \fmffreeze
                     \fmf{phantom}{oun1,oun1x1,oun1x2,oun1x3,oun1x4,oun1x5,oun1x6,oun1x7,oun1x8,oun1x9,oun1x10,oun2}
                     \fmffreeze
                     \fmf{plain,tension=0.2,left=0.3,label=$s_{n-5}$}{oun1x1,oun1x10}
                     \fmffreeze
                     \fmf{phantom}{ou5,ou5x1,ou5x2,ou5x3,ou5x4,ou5x5,ou5x6,ou5x7,ou5x8,ou5x9,ou5x10,ou4}
                     \fmffreeze
                     \fmf{plain,tension=0.2,left=0.3,label=$s_2$}{ou5x1,ou5x10}
                     \fmffreeze
                     \fmf{phantom}{ou4,ou4x1,ou4x2,ou4x3,ou4x4,ou4x5,ou4x6,ou4x7,ou4x8,ou4x9,ou4x10,u3}
                     \fmffreeze
                     \fmf{plain,tension=0.2,left=0.3,label=$s_1$}{ou4x1,ou4x10}
                     \fmffreeze
                     \fmf{phantom}{v23,v23x1,v23x2,v23x3,v23x4,v23x5,v23x6,v4}
                     \fmf{phantom}{p3,p3x1,p3x2,p3x3,p3x4,p3x5,p3x6,p4}
                     \fmffreeze
                     \fmf{gluon,label=$p_4$,l.s=left}{p3x3,v23x1}
                     \fmfv{label=$\kappa_{n-5}$,l.a=-180,l.d=-0.17w}{p4}
                     \fmfv{label=$\kappa_{n-6}$,l.a=-180,l.d=-0.17w}{p5}
                     \fmfv{label=$\kappa_{2}$,l.a=-180,l.d=-0.09w}{pn2}
                     \fmfv{label=$\kappa_{1}$,l.a=-180,l.d=-0.09w}{pn1}
                     \fmf{phantom}{v5,v67,vn2}
                     \fmfv{label=$\vdots$,l.d=-1mm}{v67}
                           \end{fmfgraph*}}
(b) \parbox{10mm}{ \begin{fmfgraph*}(150,250)
                  \fmfstraight
                     \fmfleft{p1,pdn1,pdn2,pd5,pd4,p2}
                     \fmfright{x1,x2,x3,x5,x6,x7}
                     \fmf{phantom}{p1,u1,v1n,u2,pn,u3,x1}
                     \fmf{phantom}{p2,o1,v23,o2,p3,o3,x7}
                     \fmffreeze
                     \fmf{phantom}{pn,pn1,pn2,p5,p4,p3}
                     \fmffreeze
                     \fmf{gluon,label=$p_2$,label.side=left,l.d=0.03w}{p2,v23}
                     \fmf{gluon,label=$p_3$,label.side=left,l.d=0.03w}{v23,p3}
                     \fmf{phantom}{v23,v4,v5,vn2,vn1,v1n}
                     \fmffreeze
                     \fmfv{decor.shape=circle,decor.filled=shaded,decor.size=0.19w,fore=green}{v23}
                     \fmfv{decor.shape=circle,decor.filled=shaded,decor.size=.09w,fore=green}{v4}
                     \fmfv{decor.shape=circle,decor.filled=shaded,decor.size=.09w,fore=green}{v5}
                     \fmfv{decor.shape=circle,decor.filled=shaded,decor.size=.09w,fore=green}{vn2}
                     \fmfv{decor.shape=circle,decor.filled=shaded,decor.size=.09w,fore=green}{vn1}
                     \fmfv{decor.shape=circle,decor.filled=shaded,decor.size=.19w,fore=green}{v1n}
                     \fmf{zigzag,label=$q_{n-5}$,label.side=right,l.d=0.055w}{v23,v4}
                     \fmf{zigzag,label=$q_{n-6}$,label.side=right,l.d=0.055w}{v4,v5}
                     \fmf{zigzag,label=$q_{2}$,label.side=right,l.d=0.055w}{vn2,vn1}
                     \fmf{zigzag,label=$q_{1}$,label.side=right,l.d=0.055w}{vn1,v1n}
                     \fmf{gluon,label=$p_1$,label.side=right,l.d=0.055w}{p1,v1n}
                     \fmf{gluon,label=$p_{n}$,label.side=right,l.d=0.055w}{v1n,pn}
                     \fmffreeze
                     \fmf{gluon,label=$p_5$,label.side=left,l.d=0.03w}{v4,p4}
                     \fmf{gluon,label=$p_6$,label.side=left,l.d=0.03w}{v5,p5}
                     \fmf{gluon,label=$p_{n-3}$,label.side=left,l.d=0.03w}{vn2,pn2}
                     \fmf{gluon,label=$p_{n-2}$,label.side=left,l.d=0.03w}{vn1,pn1}
                     \fmffreeze
                     \fmf{phantom}{u3,ou4,ou5,oun2,oun1,o3}
                     \fmffreeze
                     \fmf{phantom}{o3,ox1,ox2,ox3,ox4,ox5,ox6,ox7,ox8,ox9,ox10,oun1}
                     \fmffreeze
                     \fmf{plain,tension=0.2,left=0.3,label=$s_{n-5}$}{ox4,ox10}
                     \fmffreeze
                     \fmf{phantom}{oun1,oun1x1,oun1x2,oun1x3,oun1x4,oun1x5,oun1x6,oun1x7,oun1x8,oun1x9,oun1x10,oun2}
                     \fmffreeze
                     \fmf{plain,tension=0.2,left=0.3,label=$s_{n-6}$}{oun1x1,oun1x10}
                     \fmffreeze
                     \fmf{phantom}{ou5,ou5x1,ou5x2,ou5x3,ou5x4,ou5x5,ou5x6,ou5x7,ou5x8,ou5x9,ou5x10,ou4}
                     \fmffreeze
                     \fmf{plain,tension=0.2,left=0.3,label=$s_2$}{ou5x1,ou5x10}
                     \fmffreeze
                     \fmf{phantom}{ou4,ou4x1,ou4x2,ou4x3,ou4x4,ou4x5,ou4x6,ou4x7,ou4x8,ou4x9,ou4x10,u3}
                     \fmffreeze
                     \fmf{plain,tension=0.2,left=0.3,label=$s_1$}{ou4x1,ou4x7}
                     \fmffreeze
                     \fmf{phantom}{v23,v23x1,v23x2,v23x3,v23x4,v23x5,v23x6,v4}
                     \fmf{phantom}{p3,p3x1,p3x2,p3x3,p3x4,p3x5,p3x6,p4}
                     \fmf{phantom}{v1n,v1nx1,v1nx2,v1nx3,v1nx4,v1nx5,v1nx6,vn1}
                     \fmf{phantom}{pn,pnx1,pnx2,pnx3,pnx4,pnx5,pnx6,pn1}
                     \fmffreeze
                     \fmf{gluon,label=$p_4$,l.s=left}{p3x3,v23x1}
                     \fmf{gluon,label=$p_{n-1}$,l.s=right,l.d=0.07w}{pnx2,v1n}
                     \fmfv{label=$\kappa_{n-6}$,l.a=-180,l.d=-0.17w}{p4}
                     \fmfv{label=$\kappa_{n-7}$,l.a=-180,l.d=-0.17w}{p5}
                     \fmfv{label=$\kappa_{2}$,l.a=-180,l.d=-0.09w}{pn2}
                     \fmfv{label=$\kappa_{1}$,l.a=-180,l.d=-0.09w}{pn1}
                     \fmf{phantom}{v5,v67,vn2}
                     \fmfv{label=$\vdots$,l.d=-1mm}{v67}
                           \end{fmfgraph*}}
                           \vskip 1cm
                             \end{fmffile}
                            
\caption{\label{fig:quasiMR}Amplitudes in the quasi-multi-Regge kinematics of (a) a pair at either end of the ladder and (b) two pairs,
one at each end of the ladder.}
\end{figure}
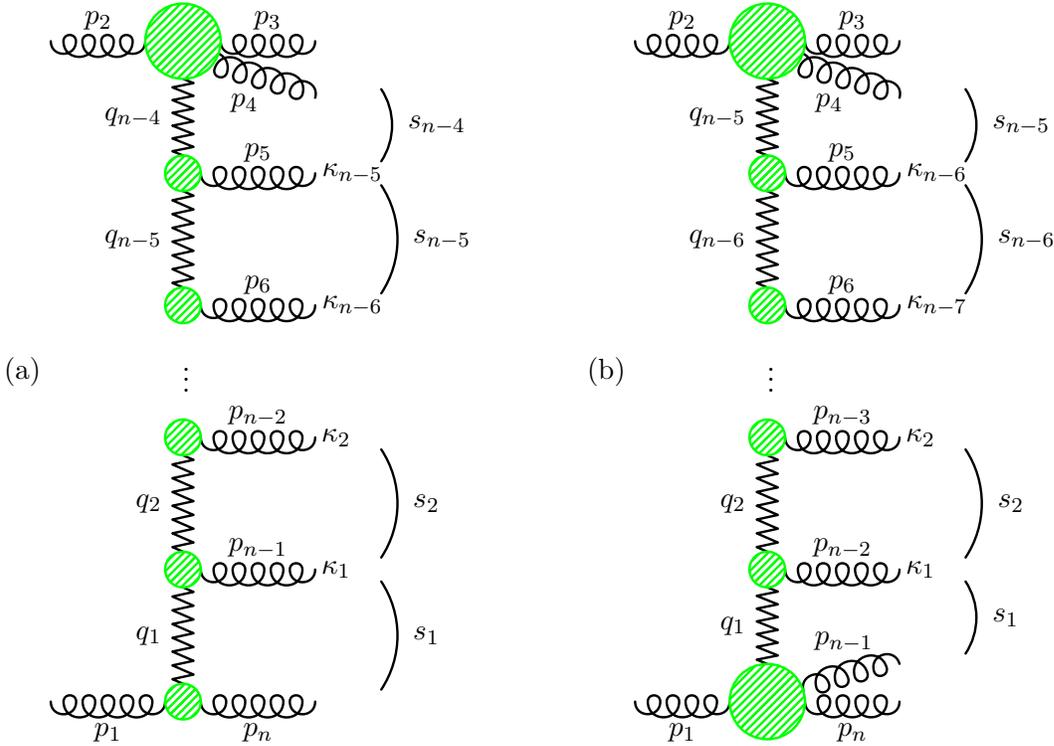

It is possible to define a high-energy prescription
for more general, {\it i.e.} less restrictive, multi-Regge kinematics, such as the
quasi-multi-Regge kinematics where all gluons are strongly
ordered in rapidity, except for a pair of gluons, either at the top or at the bottom of the ladder
as shown schematically in Fig.~\ref{fig:quasiMR}(a).
For example,
\begin{equation}
y_3 \simeq y_4\gg \cdots\gg y_n;\qquad |p_{3\perp}| \simeq |p_{4\perp}| ...\simeq|p_{n\perp}|\, ,
\label{qmrknpt}
\end{equation}
for which the Mandelstam invariants are given in \sec{sec:nptampqmr}.
We conjecture that in the quasi-multi-Regge kinematics of \eqn{qmrknpt}
a generic colour-stripped $l$-loop $n$-gluon amplitude will have the factorised form,
\bea
\lefteqn{ m_n(1,2, \ldots ,n) =
s \left[g^2\, A(p_2,p_3,p_4) \right]\, 
{1\over t_{n-4}}\, \left({-s_{n-4}\over \tau}\right)^{\alpha(t_{n-4})}
\left[g\,V(q_{n-4},q_{n-5},\kappa_{n-5})\right]  }
\nn\\ &&\qquad\qquad \cdots \times\ 
{1\over t_2}\, \left({-s_2\over \tau}\right)^{\alpha(t_2)}
\left[g\,V(q_2,q_1,\kappa_1)\right]\, 
{1\over t_1}\, \left({-s_1\over \tau}\right)^{\alpha(t_1)}
 \left[g\, C(p_1,p_n) \right]\, ,\label{loopnqmr}
\eea
where we suppressed the dependence of the coefficient functions and the Lipatov vertices
on the reggeisation scale $\tau$. $s_{n-4}$ can be chosen to be
either $s_{35}$ or $s_{45}$, the difference between the two being of the order of $s_{34}$,
thus sub-leading with respect to $s$.
In order for $m_n$ to be real, one can take the invariants $s$, $s_1, \ldots ,s_{n-4}$,
$t_1, \ldots, t_{n-4}$, defined as in \sec{sec:mrkforall}, and $s_{34}$ all negative.
Then the kinematics imply
\beq
- s \gg -s_1, -s_2, \ldots, -s_{n-4} \gg -s_{34}, -t_1, -t_2, \dots, -t_{n-4}\, .\label{eq:negqmrnpt}
\eeq
The limit of multi-Regge kinematics (\ref{eq:mrkneg}) is where $s_{34}$ becomes 
as large as any $s_i$-type invariant.

In \eqn{loopnqmr}, the coefficient function $C$ and Lipatov vertex $V$ 
are exactly the same as in \eqn{loopnpt}. However a new coefficient function,
$A(p_2,p_3,p_4)$, is needed to describe the production of two gluons at one end of the ladder.
The tree approximation, $A^{(0)}(p_2,p_3,p_4)$, was
computed in Ref.~ \cite{DelDuca:1995ki,Fadin:1996nw}.
$A$ can be expanded in the rescaled
coupling, just as in \eqns{fullv}{eq:coeffrescal},
\beq
A(p_2,p_3,p_4,\tau) = A^{(0)}(p_2,p_3,p_4)\left(1 + \tgs^{2} \bA^{(1)}(t,s_{34},\tau) 
+ \tgs^4 \bA^{(2)}(t,s_{34},\tau) + \ord(\tgs^{6}) \right)\, .\label{avert}
\eeq
For $n=5$, \eqn{loopnqmr} reduces to 
\beq
m_5(1,2,3,4,5) = s \left[g^2\, A(p_2,p_3,p_4,\tau) \right]\, 
{1\over t}\, \left({-s_1\over \tau}\right)^{\alpha(t)} \left[g\, C(p_1,p_5,\tau) \right]\, ,\label{loop5qmr}
\eeq
with $q=p_1+p_5=-(p_2+p_3+p_4)$, $t=q^2$ and $s=s_{12}$. 
Expanding \eqn{loop5qmr} as in \eqn{elasexpand}, we obtain,
at one-, two- and three-loop accuracy,
\bea
m_5^{(1)} &=& \balpha^{(1)}(t) L + \bC^{(1)}(t,\tau) + \bA^{(1)}(t,s_{34},\tau)\, ,\label{5pt1lqmr}\\
m_5^{(2)} &=& {1\over 2} \left( m_5^{(1)} \right)^2 + \balpha^{(2)}(t) L \nn\\
&+&  \bC^{(2)}(t,\tau) + \bA^{(2)}(t,s_{34},\tau) 
- {1\over 2} \left( \bC^{(1)}(t,\tau) \right)^2
- {1\over 2} \left( \bA^{(1)}(t,s_{34},\tau) \right)^2\, ,\label{5pt2lqmr}\\
m_5^{(3)} &=& m_5^{(2)}\,m_5^{(1)}-{1\over 3}\left(m_5^{(1)}\right)^3 + \balpha^{(3)}(t) L \nn\\
&+&\bC^{(3)}(t,\tau) +\bA^{(3)}(t,s_{34},\tau)
-\bC^{(2)}(t,\tau)\bC^{(1)}(t,\tau)-\bA^{(2)}(t,s_{34},\tau)\bA^{(1)}(t,s_{34},\tau)\nn\\
&+&{1\over 3}\left(\bC^{(1)}(t,\tau)\right)^3+{1\over 3}\left(\bA^{(1)}(t,s_{34},\tau)\right)^3,\,\label{5pt3lqmr}
\eea
with $L=\ln(-s_1/\tau)$, and where $m_5^{(1)}$ is needed to $\ord(\eps^2)$ in Eq.~(\ref{5pt2lqmr}), and $m_5^{(1)}$ and $m_5^{(2)}$ to $\ord(\eps^4)$ and $\ord(\eps^2)$ respectively in Eq.~(\ref{5pt3lqmr}).
The coefficient functions $\bC$ were already evaluated
in \sec{sec:4pthel}.  Therefore, knowledge of the five-point amplitude at a given loop accuracy in the
quasi-multi-Regge kinematics (\ref{qmrknpt}) allows one to find the
coefficient function $A$ at the same loop accuracy. Furthermore, combining the iterative
formula (\ref{eq:ite2})
for $n=5$ with the high-energy prescription (\ref{loopnqmr}), one obtains an iterative
formula for the coefficient function $A$,
\beq
A^{(2)}(t,s_{34},\tau,\eps) = {1\over 2} \left[ A^{(1)}(t,s_{34},\tau,\eps)\right]^2
+ {2\,G^2(\eps)\over G(2\eps)} f^{(2)}(\eps)\, A^{(1)}(t,s_{34},\tau,2\eps)
+ 2\, Const^{(2)}
+ \ord(\eps)\, ,\label{eq:avertif}
\eeq
where the one-loop coefficient function, $A^{(1)}(\eps)$, is needed to $\ord(\eps^2)$. Similarly, it is straight forward to derive from Eq.~(\ref{5pt3lqmr}) an iterative formula at the three-loop coefficient function
\bea
A^{(3)}(t,s_{34},\tau,\eps) &=& A^{(2)}(t,s_{34},\tau,\eps)A^{(1)}(t,s_{34},\tau,\eps) -{1\over 3} \left[ A^{(1)}(t,s_{34},\tau,\eps)\right]^3\nn \\
&+& {4\,G^3(\eps)\over G(3\eps)} f^{(3)}(\eps)\, A^{(1)}(t,s_{34},\tau,3\eps)
+ 4\, Const^{(3)}
+ \ord(\eps)\, ,\label{eq:avertif3l}
\eea
where the one and two-loop coefficient functions $A^{(1)}(\eps)$ and $A^{(2)}(\eps)$ are needed to $\ord(\eps^4)$ and $\ord(\eps^2)$ respectively.

\subsection{Amplitudes in the quasi-multi-Regge kinematics with two pairs,
one at each end of the ladder}
\label{sec:ampnqmrsq}

One can also consider the
quasi-multi-Regge kinematics where all gluons are strongly
ordered in rapidity, except for two pairs of gluons, one at each end of the ladder,
\begin{equation}
y_3 \simeq y_4\gg \cdots\ \gg y_{n-1} \simeq y_n;\qquad |p_{3\perp}| \simeq |p_{4\perp}| ...\simeq|p_{n\perp}|\,  ,\label{qmrksqnpt}
\end{equation}
for which the Mandelstam invariants are given in \sec{sec:nptampqmrsq} and illustrated in Fig.~\ref{fig:quasiMR}(b).
The high-energy prescription is
\bea
\lefteqn{ m_n(1,2, \ldots ,n) =
s \left[g^2\, A(p_2,p_3,p_4) \right]\, 
{1\over t_{n-5}}\, \left({-s_{n-5}\over \tau}\right)^{\alpha(t_{n-5})}
\left[g\,V(q_{n-5},q_{n-6},\kappa_{n-6})\right]  }
\nn\\ &&\quad \cdots \times\ 
{1\over t_2}\, \left({-s_2\over \tau}\right)^{\alpha(t_2)}
\left[g\,V(q_2,q_1,\kappa_1)\right]\, 
{1\over t_1}\, \left({-s_1\over \tau}\right)^{\alpha(t_1)}
 \left[g^2\, A(p_1,p_n, p_{n-1}) \right]\, .\label{loopnqmrsq}
\eea
where we again suppressed the dependence of the coefficient functions and the Lipatov vertices
on the reggeisation scale. 
In order for $m_n$ to be real, one can take all the invariants $s$- and $t$-type to be
negative. Then the kinematics imply
\beq
- s \gg -s_1, -s_2, \ldots, -s_{n-5} \gg -s_{34}, -s_{n-1,n}, -t_1, -t_2, \dots, -t_{n-5}\, ,\label{eq:negqmrsqnpt}
\eeq
The limit of multi-Regge kinematics (\ref{eq:mrkneg}) is where $s_{34}$ and
$s_{n-1,n}$ become as large as any $s_i$-type invariant.

For $n=6$, \eqn{loopnqmrsq} reduces to two coefficient functions $A$
linked by a $t$-channel reggeised gluon propagator,
\beq
m_6(1,2,3,4,5,6) =
s \left[g^2\, A(p_2,p_3,p_4,\tau) \right]\, 
{1\over t}\, \left({-s_1\over \tau}\right)^{\alpha(t)}
 \left[g^2\, A(p_1,p_n, p_{n-1},\tau) \right]\, ,\label{loop6qmrsq}
\eeq
with $q=p_1+p_5 + p_6=-(p_2+p_3+p_4)$, $t=q^2$ and $s=s_{12}$. $s_1$ can be
anything between $s_{45}$, $s_{46}$, $s_{35}$ and $s_{36}$, the difference between them 
being of the order of $s_{34}$ or $s_{56}$, thus sub-leading with respect to $s$.
The quasi-multi-Regge kinematics (\ref{eq:negqmrnpt}) become
\beq
- s \gg -s_1 \gg -s_{34}, -s_{56}, -t\, .\label{eq:negqmr6pt}
\eeq
Expanding \eqn{loop6qmrsq} as in \eqn{elasexpand}, at one-, two- and three-loop accuracy,
we obtain
\bea
m_6^{(1)} &=& \balpha^{(1)}(t) L + \bA^{(1)}(t,s_{34},\tau) + \bA^{(1)}(t,s_{56},\tau)\, ,\label{6pt1lqmr}\\
m_6^{(2)} &=& {1\over 2} \left( m_6^{(1)} \right)^2 + \balpha^{(2)}(t) L \label{6pt2lqmr}\\
&+& \bA^{(2)}(t,s_{34},\tau) + \bA^{(2)}(t,s_{56},\tau)  
- {1\over 2} \left( \bA^{(1)}(t,s_{34},\tau) \right)^2
- {1\over 2} \left( \bA^{(1)}(t,s_{56},\tau) \right)^2
\, ,\nn\\
m_6^{(3)} &=& m_6^{(2)}\,m_6^{(1)}-{1\over 3}\left(m_6^{(1)}\right)^3 + \balpha^{(3)}(t) L\label{6pt3lqmr}\nn\\
&+&\bA^{(3)}(t,s_{34},\tau) +\bA^{(3)}(t,s_{56},\tau)\nn\\
&-&\bA^{(2)}(t,s_{34},\tau)\bA^{(1)}(t,s_{34},\tau)-\bA^{(2)}(t,s_{56}\tau)\bA^{(1)}(t,s_{56},\tau)\nn\\
&+&{1\over 3}\left(\bA^{(1)}(t,s_{34},\tau)\right)^3+{1\over 3}\left(\bA^{(1)}(t,s_{56},\tau)\right)^3,\,
\eea
with $L=\ln(-s_1/\tau)$.
In the two- and three-loop expansion of the six-point amplitude,~(\ref{6pt2lqmr}) and~(\ref{6pt3lqmr}), 
no new vertices or coefficient functions occur. Thus, using the explicit expressions
of $A^{(k)}$ and $\alpha^{(k)}$, $k=1,2,3$, in \eqn{6pt2lqmr} and in \eqn{6pt3lqmr}, 
one can assemble the two- and three-loop six-point amplitude in the quasi-multi-Regge kinematics
(\ref{eq:negqmr6pt}).
However, even without knowing the explicit expression of $A^{(1)}$ and $A^{(2)}$,
it is easy to see by substitution that 
the iterative structure of Eqs.~(\ref{eq:alphabds})
and (\ref{eq:avertif}) ensures that the six-point amplitude~(\ref{6pt2lqmr})
fulfils the two-loop iterative formula (\ref{eq:ite2}) for $n=6$. 
Similarly, using \eqn{eq:avertif3l}, one can easily show that the six-point amplitude~(\ref{6pt3lqmr})
fulfils the three-loop iterative formula (\ref{eq:ite3}) for $n=6$.
Thus, also for the quasi-multi-Regge kinematics of \eqn{eq:negqmr6pt}
the quantities $R_6^{(2)}$ and $R_6^{(3)}$ vanish.

Because no new vertices or coefficient functions occur in the two- and three-loop expansion
of \eqn{loopnqmrsq} for $n > 6$, we conclude that the two- and three-loop expansions
of \eqn{loopnqmrsq} fulfil the two- and three-loop iterative formulas (\ref{eq:ite2}) and (\ref{eq:ite3}). Furthermore, it is straightforward to extend the proof of Section~\ref{sec:proof} to the  kinematics with a pair of gluons emitted at either side or at each end of the ladder, and hence the BDS
ansatz is fulfilled in quasi-multi-Regge kinematics for any $n$ or, in other words, the quantities $R_n^{(l)}$ vanish in the quasi-multi-Regge kinematics (\ref{qmrksqnpt}), for any $n$ and for any $l$.

Continuing the kinematics (\ref{eq:negqmr6pt}) to the physical region
where $s,\, s_1,\, s_{34},\, s_{56}$ are positive and $t$ is negative, 
the conformal invariants (\ref{thrinvar}) become~\cite{Brower:2008nm}
\beq
u_1  \simeq 1\, ,\quad 
u_2 \simeq \frac{(|p_{3\perp}|^2 + p_4^+p_3^-)\, s_{56}}{|q_\perp|^2\, (s_{45}+s_{46})} 
\simeq \ord\left(\frac{t}{s}\right)\, ,\quad
u_3 = \frac{(|p_{6\perp}|^2 + p_6^+p_5^-)\, s_{34} }{|q_\perp|^2\, (s_{35}+s_{45})} 
\simeq \ord\left(\frac{t}{s}\right)\, ,\label{thrinvarqmrk}
\eeq
thus, just like for the multi-Regge kinematics (\ref{eq:mrk6pt})
$u_1$ is close to 1, while $u_2$ and $u_3$ are very small, in fact sub-leading
to the desired accuracy.

\section{What lies beyond?}
\label{sec:outlook}

From the analysis of Sects.~\ref{sec:bdsmrk} and \ref{sec:quasi}, it is
clear that no difference between the Regge factorisation and the BDS ansatz
will be found, unless there is a contribution from
coefficient funtions which appear for the first time 
in $n$-gluon amplitudes, with $n\ge 6$.
To introduce this type of coefficient function means considering even less restrictive
multi-Regge kinematics.
In this Section, we examine the two simplest of such instances: a cluster of
two gluons along the ladder, and a cluster of three gluons at one end of the
ladder.

\subsection{Six-point amplitude in the quasi-multi-Regge kinematics of a pair
along the ladder}
\label{sec:ampnqmrc}

\begin{figure}[!t]
 \begin{fmffile}{sixpoints}
   \qquad\quad(a) \parbox{60mm}{\begin{fmfgraph*}(150,115)
                  \fmfstraight
                     \fmfleft{p1,p2}
                     \fmfright{x1,x2}
                     \fmf{phantom}{p1,u1,v16,u2,p6,x1}
                     \fmf{phantom}{p2,o1,v23,o2,p3,x2}
                     \fmf{phantom}{p6,y1,xy1,x1}
                     \fmf{phantom}{p3,y2,xy2,x2}
                     \fmffreeze
                     \fmf{phantom}{v16,vx1,vx2,vc,vx3,x4,v23}
                     \fmf{phantom}{p3,px1,px2,p4,pc,p5,px3,px4,p6}
                     \fmf{phantom}{x1,xx1,xx2,xx3,xc,xx4,xx5,xx6,x2}
                     \fmf{phantom}{y1,yy1,yy2,yy3,yc,yy4,yy5,yy6,y2}
                     \fmffreeze
                     \fmf{zigzag,label=$q_2$,label.side=right,l.d=0.05w}{v23,vc}
                     \fmf{zigzag,label=$q_1$,label.side=right,l.d=0.05w}{vc,v16}
                     \fmf{gluon,label=$p_2$,label.side=left,l.d=0.03w}{p2,v23}
                     \fmf{gluon,label=$p_3$,label.side=left,l.d=0.03w}{v23,p3}
                     \fmf{gluon,label=$p_1$,label.side=left,l.d=0.03w}{p1,v16}
                     \fmf{gluon,label=$p_6$,label.side=left,l.d=0.03w}{v16,p6}
                     \fmffreeze
                     \fmfv{decor.shape=circle,decor.filled=shaded,decor.size=0.09w,fore=green}{v23}
                     \fmfv{decor.shape=circle,decor.filled=shaded,decor.size=.09w,fore=green}{v16}
                     \fmfv{decor.shape=circle,decor.filled=shaded,decor.size=.21w,fore=green}{vc}
                     \fmffreeze
                     \fmf{gluon,label=$p_4$}{vc,p4}
                     \fmf{gluon,label=$p_5$,l.d=0.06w,label.side=right}{vc,p5}
                     \fmf{plain,tension=0.2,right=0.3,label=$s_1$}{y1,yy3}
                     \fmf{plain,tension=0.2,left=0.3,label=$s_2$}{y2,yy4}
                     \end{fmfgraph*}}
                   (b)   \parbox{20mm}{\begin{fmfgraph*}(150,120)
                  \fmfstraight
                     \fmfleft{p1,p2}
                     \fmfright{p6,p3}
                     \fmf{phantom}{p1,v16,p6}
                     \fmf{phantom}{p2,v23,p3}
                     \fmf{phantom}{p6,aaa,p3}
                     \fmf{phantom}{p3,p4,p5,aaa}
                     \fmffreeze
                     \fmf{phantom}{v23,bbb,v16}
                     \fmf{phantom}{bbb,u1,u2,u3,u4,u5,v23}
                      \fmf{phantom}{bbb,uu1,uu2,uu3,uu4,v16}
                      \fmffreeze
                     \fmf{phantom,label=$q$,label.side=right,l.d=0.05w}{bbb,uu3}
                     \fmffreeze
                     \fmf{zigzag}{v16,v23}
                     \fmf{gluon,label=$p_1$,label.side=left,l.d=0.03w}{p1,v16}
                     \fmf{gluon,label=$p_6$,label.side=left,l.d=0.03w}{v16,p6}
                     \fmf{gluon,label=$p_2$,label.side=left,l.d=0.03w}{p2,v23}
                     \fmf{gluon,label=$p_3$,label.side=left,l.d=0.03w}{v23,p3}
                      \fmf{gluon}{u5,p4}
                       \fmf{gluon,label=$p_5$,label.side=right,l.d=0.06w}{u4,p5}
                     \fmffreeze
                     \fmfv{label=$p_4$,l.a=-10}{p4}
                     \fmfv{decor.shape=circle,decor.filled=shaded,decor.size=.09w,fore=green}{v16}
                     \fmfv{decor.shape=circle,decor.filled=shaded,decor.size=.30w,fore=green}{u5}
                     \end{fmfgraph*}}
                     \vskip 1cm
                                                        \end{fmffile}
\caption{\label{fig:6point}Six-point amplitude in the quasi-multi-Regge kinematics of (a) a pair along the ladder and (b) three-of-a-kind.}
\end{figure}
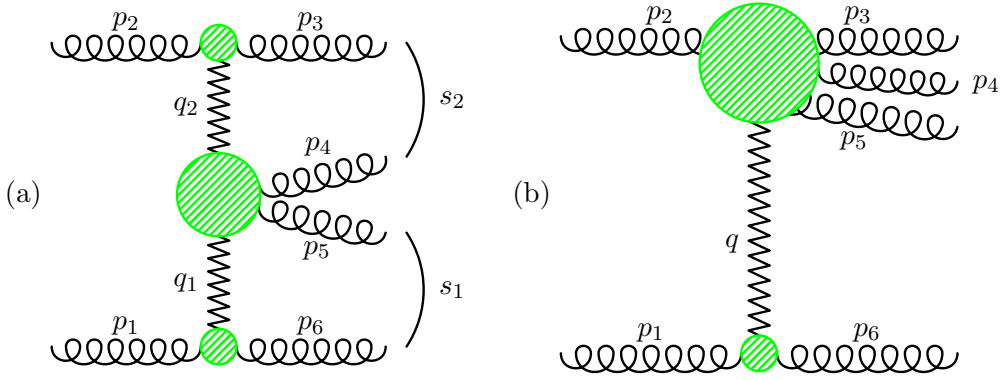

In the quasi-multi-Regge kinematics of \sec{sec:nptampqmrc}, where the outgoing
gluons are strongly ordered in rapidity, except for the central pair,
\begin{equation}
y_3 \gg y_4 \simeq y_5 \gg y_6;\qquad |p_{3\perp}| \simeq |p_{4\perp}| 
\simeq |p_{5\perp}| \simeq|p_{6\perp}|\, 
,\label{qmrk6ptc}
\end{equation}
the high-energy prescription is
\bea
m_6(1,2,3,4,5,6) &=& s \left[g\, C(p_2,p_3,\tau) \right]\, 
{1\over t_2}\, \left({-s_2\over \tau}\right)^{\alpha(t_2)} \nn\\
&\times& \left[g^2\,W(q_2,q_1,p_4,p_5,\tau)\right]\, 
{1\over t_1}\, \left({-s_1\over \tau}\right)^{\alpha(t_1)}
 \left[g\, C(p_1,p_6,\tau) \right]\, , \label{2lLip6pt}
\eea
where $p_4+p_5=q_2-q_1$, and with 
$t_1=s_{61}$, $t_2=s_{23}$, $s_1=s_{56}$ and $s_2=s_{34}$ as illustrated in~\fig{fig:6point}(a).
In order for the amplitude $m_6$ to be real,
\eqn{2lLip6pt} is taken in the region where all the invariants are negative.
Thus, the quasi-multi-Regge kinematics (\ref{qmrk6ptc}) become,
\begin{equation}
-s \gg -s_{1}, -s_{2} \gg -s_{45}, -t_1, -t_2\, .\label{eq:mrk2lLip} 
\end{equation}
In \eqn{2lLip6pt} a new coefficient function occurs: the vertex for the emission
of two gluons along the ladder, 
$W(q_2,q_1,p_4,p_5,\tau)$, which we shall call the two-gluon Lipatov vertex.
Although \eqn{2lLip6pt} can be defined for a generic helicity configuration,
the MHV amplitude requires the two-gluon Lipatov vertex to have two gluons
of equal helicity. $W$ can be expanded in the rescaled coupling, 
\bea
W(q_2,q_1,p_4,p_5,\tau) &=& W^{(0)}(q_2,q_1,p_4,p_5) \nn\\
&\times& \left(1 + \tgs^{2} \bW^{(1)}(t_1,t_2,s_{45},\tau) 
+ \tgs^4 \bW^{(2)}(t_1,t_2,s_{45},\tau) + \ord(\tgs^{6}) \right)\, .\label{2lipvert}
\eea
The tree approximation, $W^{(0)}(q_2,q_1,p_4,p_5)$, was
computed in Ref.~\cite{DelDuca:1995ki,Fadin:1996nw}. The one-loop coefficient,
$W^{(1)}(t_1,t_2,s_{45},\tau)$ is known for the equal-helicity 
configuration~\cite{Bartels:2008ce}.
Expanding \eqn{2lLip6pt} to one-, two-, and three-loop accuracy, we obtain
\bea
m_6^{(1)} &=& \balpha^{(1)}(t_1) L_1 + \balpha^{(1)}(t_2) L_2
+ \bC^{(1)}(t_1,\tau) + \bC^{(1)}(t_2,\tau) + \bW^{(1)}(t_1,t_2,s_{45},\tau)\, \nn\\
m_6^{(2)} &=& \frac{1}{2} \left( m_6^{(1)} \right)^2
+ \balpha^{(2)}(t_1) L_1 + \balpha^{(2)}(t_2) L_2 \label{w2loop}\\
&+& \bC^{(2)}(t_1,\tau) + \bC^{(2)}(t_2,\tau) + 
\bW^{(2)}(t_1,t_2,s_{45},\tau) \nn\\
&-& \frac{1}{2} \left( \bC^{(1)}(t_1,\tau) \right)^2
- \frac{1}{2} \left( \bC^{(1)}(t_2,\tau) \right)^2
- \frac{1}{2} \left( \bW^{(1)}(t_1,t_2,s_{45},\tau) \right)^2\, ,\nn\\
m_6^{(3)} &=& m_6^{(2)}\,m_6^{(1)}-{1\over 3}\left(m_6^{(1)}\right)^3 + \balpha^{(3)}(t) L_1+ \balpha^{(3)}(t) L_2\label{w3loop}\\
&+&\bC^{(3)}(t_1,\tau) + \bC^{(3)}(t_2,\tau) +\bW^{(3)}(t_1, t_2,s_{45},\tau)\nn\\
&-&\bC^{(2)}(t_1,\tau)\bC^{(1)}(t_1,\tau)-\bC^{(2)}(t_2,\tau)\bC^{(1)}(t_2,\tau)-\bW^{(2)}(t_1,t_2,s_{45}\tau)\bW^{(1)}(t_1,t_2,s_{45},\tau)\nn\\
&+&{1\over 3}\left(\bC^{(1)}(t_1,\tau)\right)^3+{1\over 3}\left(\bC^{(1)}(t_2,\tau)\right)^3+{1\over 3}\left(\bW^{(1)}(t_1,t_2,s_{45},\tau)\right)^3,\,\nn
\eea
with $L_i=\ln(-s_i/\tau)$ and $i=1,2$, and where $m_6^{(1)}$ must be known
to $\ord(\eps^2)$ in \eqn{w2loop} and $m_6^{(1)}$ and $m_6^{(2)}$ to  $\ord(\eps^4)$ and  $\ord(\eps^2)$ respectively in \eqn{w3loop}.
Because for $n=6$ we expect to find a remainder function $R_6^{(2)}$,
combining the iterative formula (\ref{eq:discr}) with the two-loop expansion
(\ref{w2loop}), we obtain an iterative formula for the vertex $W^{(2)}$,
\bea
W^{(2)}(t_1,t_2,s_{45},\tau,\eps) &=& {1\over 2} \left[ W^{(1)}(t_1,t_2,s_{45},\tau,\eps)\right]^2 
\label{eq:wif}\\
&+& {2\,G^2(\eps)\over G(2\eps)} f^{(2)}(\eps)\, W^{(1)}(t_1,t_2,s_{45},\tau,2\eps) 
+ R_6^{(2)}(u_1^W,u_2^W,u_3^W) + \ord(\eps)\, ,\nn
\eea
where the one-loop coefficient, $W^{(1)}(\eps)$, is needed to $\ord(\eps^2)$.
Thus, a remainder function $R_6^{(2)}$ for the multi-Regge kinematics (\ref{eq:mrk2lLip})
may occur in the two-loop iteration of the two-gluon Lipatov vertex.

Using the Mandelstam invariants of \sec{sec:nptampqmrc}, the conformal invariants (\ref{thrinvar}) become
\bea
u_1&\rightarrow& u_1^W  = \frac{s_{45}}{(p_4^++p_5^+)(p_4^-+p_5^-)} \simeq \ord(1)\, , \nn\\
u_2&\rightarrow& u_2^W= \frac{|p_{3\perp}|^2  p_5^+p_6^-}{(|p_{3\perp}+p_{4\perp}|^2 + p_5^+p_4^-) 
(p_4^++p_5^+)p_6^- }\simeq \ord(1)\, , \nn\\
u_3&\rightarrow& u_3^W = \frac{|p_{6\perp}|^2  p_3^+p_4^- }{p_3^+ (p_4^-+p_5^-) 
(|p_{3\perp}+p_{4\perp}|^2 + p_5^+p_4^-) }\simeq \ord(1)\, 
,\label{thrinvarqmrkc}
\eea
{\it i.e.} all the invariants yield a non-vanishing contribution, which is in general
different from unity.

\subsection{Six-point amplitude in the quasi-multi-Regge kinematics of three-of-a-kind}
\label{sec:amp6qmr3}

In the quasi-multi-Regge kinematics of \sec{sec:kin6qmr3}, where the outgoing
gluons are emitted three in a cluster on one end and one on the other end of the ladder,
\begin{equation}
y_3 \simeq y_4 \simeq y_5 \gg y_6;\qquad 
|p_{3\perp}| \simeq |p_{4\perp}| \simeq |p_{5\perp}| \simeq|p_{\perp}|\, 
,\label{qmrk6pt3}
\end{equation}
the high-energy prescription is
\bea
m_6(1,2,3,4,5,6) &=& s \left[g\, B(p_2,p_3,p_4,p_5,\tau) \right]\, 
{1\over t}\, \left({-s_1\over \tau}\right)^{\alpha(t)}  \left[g\, C(p_1,p_6,\tau) \right]\, , \label{B3Lip6pt}
\eea
where $q=p_1+p_6$, as shown in~\fig{fig:6point}(b), $t=q^2$ and $s=s_{12}$. $s_1$ can be
anything between $s_{36}$, $s_{46}$, and $s_{56}$, the difference between them 
being of the order of $s_{345}$, thus sub-leading with respect to $s$.
In order for the amplitude $m_6$ to be real,
\eqn{B3Lip6pt} is taken in the region where all the invariants are negative.
Thus, the quasi-multi-Regge kinematics (\ref{qmrk6pt3}) become,
\begin{equation}
-s \gg -s_{1} \gg -s_{34},-s_{45},-s_{35}, -t\, .\label{eq:mrkB3Lip} 
\end{equation}
In \eqn{B3Lip6pt} a new coefficient function occurs for the emission
of three gluons at one end of the ladder occurs,
$B(p_3,p_4,p_5,\tau)$. $B$ can be expanded in the rescaled coupling, 
\begin{equation}
\begin{split}
B(p_3,p_4,p_5,&\tau) = B^{(0)}(p_3,p_4,p_5)\\
&\times \left(1 + \tgs^{2} \bar B^{(1)}(t,s_{34},s_{45},s_{35},\tau) 
+ \tgs^4 \bar B^{(2)}(t,s_{34},s_{45},s_{35},\tau) + \ord(\tgs^{6}) \right)\, .\label{B3ipvert}
\end{split}
\end{equation}
The tree approximation, $B^{(0)}(p_3,p_4,p_5)$, was
computed in Ref.~\cite{Del Duca:1999ha}. 
Expanding \eqn{B3Lip6pt} to one-, two- and three-loop accuracy, we obtain
\bea
m_6^{(1)} &=& \balpha^{(1)}(t) L
+ \bar B^{(1)}(t,s_{34},s_{45},s_{35},\tau) + \bC^{(1)}(t,\tau)\, , \nn\\
m_6^{(2)} &=& \frac{1}{2} \left( m_6^{(1)} \right)^2
+ \balpha^{(2)}(t) L \label{B2loop}
+ \bar B^{(2)}(t,s_{34},s_{45},s_{35},\tau) + \bC^{(2)}(t,\tau) \\
&-& \frac{1}{2} \left( \bar B^{(1)}(t,s_{34},s_{45},s_{35},\tau) \right)^2
- \frac{1}{2} \left( \bC^{(1)}(t,\tau) \right)^2 ,\nn\\
m_6^{(3)} &=& m_6^{(2)}m_6^{(1)}-\frac{1}{3} \left( m_6^{(1)} \right)^3
+ \balpha^{(3)}(t) L \label{B3loop}
+ \bar B^{(3)}(t,s_{34},s_{45},s_{35},\tau) + \bC^{(3)}(t,\tau) \\
&-& \bar B^{(2)}(t,s_{34},s_{45},s_{35},\tau)\bar B^{(1)}(t,s_{34},s_{45},s_{35},\tau) - \bC^{(2)}(t,\tau)\bC^{(1)}(t,\tau)\nn\\ 
&+& \frac{1}{3} \left( \bar B^{(1)}(t,s_{34},s_{45},s_{35},\tau) \right)^3
+ \frac{1}{3} \left( \bC^{(1)}(t,\tau) \right)^3,\nn
\eea
with $L=\ln(-s_1/\tau)$, and where $m_6^{(1)}$ must be known
to $\ord(\eps^2)$ in \eqn{B2loop} and $m_6^{(1)}$ and $m_6^{(2)}$ to $\ord(\eps^4)$ and to $\ord(\eps^2)$ respectively in \eqn{B2loop}. 
Because for $n=6$ we expect to find a remainder function $R_6^{(2)}$,
combining the iterative formula (\ref{eq:discr}) with the two-loop expansion
(\ref{B2loop}), we obtain an iterative formula for the vertex $B^{(2)}$,
\bea
B^{(2)}(t,s_{34},s_{45},s_{35},\tau,\eps) &=& 
{1\over 2} \left[ B^{(1)}(t,s_{34},s_{45},s_{35},\tau,\eps)\right]^2 \label{eq:avert3g}\\
&+& {2\,G^2(\eps)\over G(2\eps)} f^{(2)}(\eps)\, B^{(1)}(t,s_{34},s_{45},s_{35},\tau,2\eps)
+ 2\, Const^{(2)} \nn\\
&+& R_6^{(2)}(u_1^B,u_2^B,u_3^B)
+ \ord(\eps)\, ,\nn
\eea
where the one-loop coefficient, $B^{(1)}(\eps)$, is needed to $\ord(\eps^2)$.
Thus, a remainder function $R_6^{(2)}$ for the multi-Regge kinematics (\ref{eq:mrkB3Lip})
may occur in the two-loop iteration for the coefficient function for the emission of three gluons 
on one end of the ladder.

In the limit $y_3\gg y_4\simeq y_5$, the kinematics (\ref{qmrk6pt3}) reduce to
\eqn{qmrk6ptc} and the prescription (\ref{B3Lip6pt}) reduces to \eqn{2lLip6pt}.
Then the coefficient function $B$ factors out into the two-gluon Lipatov vertex $W$
and the coefficient function for the emission of a gluon, linked by a reggeised
propagator~\cite{Del Duca:1999ha}. Accordingly, the remainder function 
$R_6^{(2)}(u_1^B,u_2^B,u_3^B)$
in \eqn{eq:avert3g} reduces to $R_6^{(2)}(u_1^W,u_2^W,u_3^W)$ in \eqn{eq:wif}.

Using the Mandelstam invariants of \sec{sec:kin6qmr3},
the conformal invariants (\ref{thrinvar}) become~\cite{Brower:2008nm}
\bea
u_1&\rightarrow& u_1^B  = \frac{s\, s_{45}}{s_{345} (p_4^++p_5^+) p_6^-} \simeq \ord(1)\, , \nn\\
u_2&\rightarrow& u_2^B = \frac{(|p_{3\perp}|^2 + (p_4^++p_5^+) p_3^- )  p_5^+p_6^-}
{(p_4^++p_5^+)p_6^-
(|p_{3\perp}+p_{4\perp}|^2 + (p_3^-+p_4^-) p_5^+) } \simeq \ord(1)\, , \nn\\
u_3&\rightarrow& u_3^B = \frac{|p_{6\perp}|^2  s_{34} }{s_{345}
(|p_{3\perp}+p_{4\perp}|^2 + (p_3^-+p_4^-) p_5^+)
 }\simeq \ord(1)\, 
,\label{thrinvarqmrk3g}
\eea
{\it i.e.} all the invariants are of similar size.


\section{Conclusions}

In this work we investigated the high-energy limit of
a colour-stripped MHV amplitude, which
is based on the Regge factorisation of the amplitude into a ladder of coefficient
functions and vertices linked by reggeised propagators~\cite{DelDuca:2008pj}.
We showed explicitly that in the Euclidean region
two- and three-loop $n$-gluon amplitudes in multi-Regge
kinematics are fully consistent with the Bern-Dixon-Smirnov ansatz, and in \sec{sec:proof} we proved
that this result holds true at any loop accuracy. In particular, this implies
that in the Euclidean region the breakdown of the iterative structure of the two-loop amplitudes, occurring in the 
two-loop six-point amplitude, cannot be resolved by multi-Regge kinematics,
{\it i.e.} the remainder function $R_6^{(2)}$
is sub-leading in the multi-Regge kinematics.

In \sec{sec:quasi} we showed that similar conclusions can be drawn for less restrictive
multi-Regge kinematics,
namely the kinematics where all the outgoing gluons are strongly ordered in
rapidity, but for a pair of gluons either at one end or at both ends of the ladder.
By giving explicit examples for the two- and three-loop six-point amplitude, we argued
that in this case as well the Regge factorisation of the amplitude is consistent
with the iterative structure implied by the BDS ansatz.    The structure of the high energy prescription ensures that
this result is valid for an arbitrary number of loops.

Finally, in order to find kinematics which might shed light on the 
 violation of the
BDS ansatz for the two-loop six-point amplitude,  in \sec{sec:outlook} we considered  kinematics which occur only for 
$n$-gluon amplitudes with $n\ge 6$, and thus for which we could not invoke the 
BDS iterative structure. We showed that the iterative structures for the new two-loop functions that appear in these kinematics 
might have a dependence on the remainder function $R_6^{(2)}(u_1,u_2,u_3)$, where $u_1$, $u_2$, $u_3$ are the conformal invariants, and therefore we argued that these kinematical limits could provide some information on
this quantity.
This suggestion is supported by the observation that, while in the multi-Regge kinematics of 
\sec{sec:bdsmrk} and in the quasi-multi-Regge
kinematics of \sec{sec:quasi} the three conformal cross ratios (\ref{thrinvar})
all took limiting values, in the more general quasi-multi-Regge kinematics of
\sec{sec:outlook} they are allowed to vary over a range defined by the kinematic
invariants.

\section*{Acknowledgements}

We thank Lance Dixon, Vladimir Smirnov and Gabriele Travaglini for useful discussions.
CD thanks the IPPP Durham and the LNF Frascati for
the warm hospitality at various stages of this work. CD is a research fellow of the \emph{Fonds National de la Recherche Scientifique}, Belgium. This work was partly supported by MIUR under contract 2006020509$_0$04,
and by the EC Marie-Curie Research Training Network ``Tools and Precision Calculations for Physics Discoveries at Colliders'' under contract MRTN-CT-2006-035505. EWNG gratefully acknowledges the support of the Wolfson Foundation and the Royal Society.

\section*{Erratum}
 
We also would like to thank Lance Dixon and Jochen Bartels for pointing out to us that the factorised form conjectured in Eq.~(3.4) is not valid in
the Minkowski region where the centre-of-mass energy squared $s$ and the energy squared $s_2$ 
of the two gluons emitted along the ladder are time-like while all other invariants stay space-like. 
Eq.~(3.4) is valid in the Euclidean region, where all invariants are
space-like, and in the physical region, where the $s$-type invariants are time-like and the $t$-type invariants are space-like.
This error is corrected in the present version, where we have made it clear that we are referring to the Euclidean and to the physical regions only. The non commuting of the high-energy limit and the $\eps$ expansion described in Appendix C of the previous version
is no longer relevant to the discussion, and Appendix C has been removed.

\appendix

\section{Multi-parton kinematics}
\label{sec:mpk}

We consider the production of $n-2$ gluons of momentum $p_i$, with 
$i=3,...,n$ in the scattering between two partons of  
momenta $p_1$ and $p_2$\footnote{By convention
we consider the scattering in the unphysical region where all momenta 
are taken as outgoing, and then we analitically continue to the
physical region where $p_1^0<0$ and $p_2^0<0$.}.

Using light-cone coordinates $p^{\pm}= p_0\pm p_z $, and
complex transverse coordinates $p_{\perp} = p^x + i p^y$, with scalar
product $2 p\cdot q = p^+q^- + p^-q^+ - p_{\perp} q^*_{\perp} - p^*_{\perp} 
q_{\perp}$, the 4-momenta are,
\begin{eqnarray}
p_2 &=& \left(p_2^+/2, 0, 0,p_2^+/2 \right) 
     =  \left(p_2^+ , 0; 0, 0 \right)\, ,\nonumber \\
p_1 &=& \left(p_1^-/2, 0, 0,-p_1^-/2 \right) 
     =  \left(0, p_1^-; 0, 0\right)\, ,\label{in}\\
p_i &=& \left( (p_i^+ + p_i^- )/2, 
                {\rm Re}[p_{i\perp}],
                {\rm Im}[p_{i\perp}], 
                (p_i^+ - p_i^- )/2 \right)\nonumber\\
    &=& \left(|p_{i\perp}| e^{y_i}, |p_{i\perp}| e^{-y_i}; 
|p_{i\perp}|\cos{\phi_i}, |p_{i\perp}|\sin{\phi_i}\right)\, \,,\nonumber
\end{eqnarray}
where $y$ is the rapidity. The first notation above is the 
standard representation 
$p^\mu =(p^0,p^x,p^y,p^z)$, while in the second we have the + and -
components on the left of the semicolon, 
and on the right the transverse components.
In the following, if not  differently stated, $p_i$ and $p_j$ are always 
understood to lie in the range $3\le i,j \le n$. The mass-shell condition is
$|p_{i\perp}|^2 = p_i^+ p_i^-$. From the momentum conservation,
\begin{eqnarray}
0 &=& \sum_{i=3}^n p_{i\perp}\, ,\nonumber \\
p_2^+ &=& -\sum_{i=3}^n p_i^+\, ,\label{nkin}\\ 
p_1^- &=& -\sum_{i=3}^n p_i^-\, ,\nonumber
\end{eqnarray}
the Mandelstam invariants may be written as,
\begin{eqnarray}
s_{ij} &=& 2 p_i\cdot p_j = p_i^+ p_j^- + p_i^- p_j^+
- p_{i\perp} p_{j\perp}^* - p_{i\perp}^* p_{j\perp}\, ,\label{eq:mandelst}
\end{eqnarray}
so that 
\begin{eqnarray}
s &=& 2 p_1\cdot p_2 = \sum_{i,j=3}^n p_i^+ p_j^-\, , \nonumber\\ 
s_{2i} &=& 2 p_2\cdot p_i = -\sum_{j=3}^n p_i^- p_j^+\, , \label{inv}\\ 
s_{1i} &=& 2 p_1\cdot p_i = -\sum_{j=3}^n p_i^+ p_j^-\, . \nonumber
\end{eqnarray}

Using the spinor representation of Ref.~\cite{Del Duca:1999ha},
\begin{equation} \begin{array}{cc}
\psi_+(p_i) = \left( \begin{array}{c} \sqrt{p_i^+}\\ \sqrt{p_i^-} e^{i\phi_i}\\ 0\\ 0\end{array}\right)\, ,
& \psi_-(p_i) = \left(
\begin{array}{c} 0\\ 0\\ \sqrt{p_i^-} e^{-i\phi_i}\\ 
-\sqrt{p_i^+}\end{array}\right)\, ,
\\ \\ \psi_+(p_2) = i \left( \begin{array}{c} \sqrt{-p_2^+}\\ 0\\
0\\ 0\end{array} \right)\, , & \psi_-(p_2) = i \left( \begin{array}{c}
0\\ 0\\ 0\\ -\sqrt{-p_2^+} \end{array}\right)\, ,\\ \\ 
\psi_+(p_1) = -i
\left( \begin{array}{c} 0\\ \sqrt{-p_1^-} \\ 0\\ 0\end{array}\right)\, , &
\psi_-(p_1) = -i \, \left( \begin{array}{c} 0\\ 0\\
\sqrt{-p_1^-}\\ 0\end{array}\right)\, . \end{array}\label{spin}
\end{equation}
for the momenta (\ref{in})\footnote{The spinors of the 
incoming partons must be continued
to negative energy after the complex conjugation, \emph{e.g.}
$\overline{\psi_{+}(p_2)}= i \left( \sqrt{-p_2^+}, 0, 0, 0 \right)$.},
the spinor products are
\begin{eqnarray}
\langle 2 1\rangle 
&=& -\sqrt{s}\, ,\nonumber\\
\langle 2 i\rangle &=& - i \sqrt{\frac{-p_2^+}{p_i^+}}\, p_{i\perp}\, ,\label{spro}\\ 
\langle i 1\rangle &=& i \sqrt{-p_1^- p_i^+}\, ,\nonumber\\ 
\langle i j\rangle &=& p_{i\perp}\sqrt{\frac{p_j^+}{p_i^+}} - p_{j\perp}
\sqrt{\frac{p_i^+}{ p_j^+}}\, , \nonumber
\end{eqnarray}
where we have used the mass-shell condition $|p_{i\perp}|^2 = p_i^+ p_i^-$.
The spinor products fulfill the usual identities,
\begin{eqnarray}
\langle i j\rangle &=& - \langle j i\rangle \nonumber\\
           \left[ i j \right] &=& - \left[ j i \right] \nonumber\\
\langle i j\rangle^* &=& {\rm sign}(p^0_i p^0_j) \left[ j i\right] 
\nonumber\\
\left( \langle i+| \gamma^{\mu}  |j+\rangle \right)^* &=&
{\rm sign}(p^0_i p^0_j)
\langle j+| \gamma^{\mu}  |i+\rangle \nonumber\\
\langle i j \rangle \left[ji\right] &=& 
2p_i\cdot p_j = \hat{s}_{ij}\,  \label{flip0}\\
\langle i+| \slash \!\!\! k  |j+\rangle  &=&
\left[ i k \right] \langle k j \rangle \nonumber \\
\langle i-| \slash  \!\!\! k  |j-\rangle  &=& 
\langle i k \rangle \left[k j\right]\,\nonumber \\
\langle ij \rangle\langle kl \rangle   &=& 
\langle ik \rangle\langle jl \rangle + 
\langle il \rangle\langle kj \rangle\,\nonumber
\end{eqnarray}
and if $\sum_{i=1}^{n} p_i=0$ then
\begin{eqnarray}
\sum_{i=1}^{n} \left[ji\right] \langle ik \rangle =0\,.
\end{eqnarray}

\section{Multi-Regge kinematics}
\label{sec:mrk}

In the multi-Regge kinematics, we require that the gluons
are strongly ordered in rapidity and have comparable transverse momentum
(\ref{mrknpt}).
This is equivalent to require a strong ordering of the light-cone coordinates,
\begin{equation}
p_3^+\gg p_4^+\cdots\gg p_n^+; \qquad p_3^-\ll p_4^-\cdots\ll p_n^-.
\end{equation}
In the high-energy limit, momentum conservation (\ref{nkin}) then becomes
\begin{eqnarray}
0 &=& \sum_{i=3}^n p_{i\perp}\, ,\nonumber \\
p_2^+ &=& -p_3^+\, ,\label{mrkin}\\ 
p_1^- &=& -p_n^-\, ,\nonumber
\end{eqnarray}
where the $=$ sign is understood to mean ``equals up to corrections of next-to-leading accuracy''.
The Mandelstam invariants (\ref{inv}) are reduced to,
\begin{eqnarray}
s &=& 2 p_1\cdot p_2 = p_3^+ p_n^-, \nonumber\\ 
s_{2i} &=& 2 p_2\cdot p_i = - p_3^+ p_i^-, \label{mrinv}\\ 
s_{1i} &=& 2 p_1\cdot p_i = - p_i^+ p_n^-, \nonumber\\ 
s_{ij} &=& 2 p_i\cdot p_j = p_i^+ p_j^-\qquad i < j\, .\nonumber
\end{eqnarray}
The product of two successive invariants of type $s_{ij}$ fixes the mass shell.
For example, 
\beq
s_{k-1,k}s_{k,k+1} = p_{k-1}^+ p_k^- p_k^+ p_{k+1}^- = 
|p_{k\perp}|^2 p_{k-1}^+ p_{k+1}^- = |p_{k\perp}|^2 s_{k-1,k+1} = |p_{k\perp}|^2 
s_{k-1,k,k+1}\, .\nn
\eeq
Thus,
\beq
|p_{k\perp}|^2 = \frac{s_{k-1,k}s_{k,k+1}}{s_{k-1,k,k+1}}\, .\label{eq:masshell}
\eeq
The spinor products (\ref{spro}) are,
\begin{eqnarray}
\langle 2 1\rangle &=& -\sqrt{p_3^+ p_n^-}\, ,\nonumber\\
\langle 2 i\rangle &=& - i\sqrt{\frac{p_3^+}{ p_i^+}}\, p_{i\perp}\, ,\label{mrpro}\\
 \langle i 1\rangle &=& i\sqrt{p_i^+ p_n^-}\, ,\nonumber\\ 
\langle i j\rangle &=& -\sqrt{\frac{p_i^+}{p_j^+}}\,
p_{j\perp}\, \qquad {\rm for}\, y_i>y_j \, .\nonumber
\end{eqnarray}

\subsection{6-point amplitude in multi-Regge kinematics}
\label{sec:6ptmrkinvar}

For $n = 6$, the momenta of the gluons exchanged in the $t$ channel
are $q_1 = p_1+p_6$, $q_2= q_1+p_5=q_3-p_4$, $q_3=-p_2-p_3$.
The cyclic Mandelstam invariants are 
\begin{eqnarray}
s &=& p_3^+ p_6^-, \nonumber\\
s_{23} &=& - p_3^+ p_3^- = - |p_{3\perp}|^2 = - |q_{3\perp}|^2, \nonumber\\ 
s_{34} &=& p_3^+ p_4^-, \nonumber\\
s_{45} &=& p_4^+ p_5^-, \nonumber\\
s_{56} &=& p_5^+ p_6^-, \nonumber\\
s_{61} &=& - p_6^+ p_6^- = - |p_{6\perp}|^2 = - |q_{1\perp}|^2.
\end{eqnarray}
Then we see that
\begin{equation}
s_{345}\, s_{456} = s\, s_{45}. \label{srel}
\end{equation}
The mass-shell conditions for the gluons along the ladder imply that
\begin{eqnarray}
s_{34}\, s_{45} &=& s_{345}\, |p_{4\perp}|^2 = s_{345}\, |q_{3\perp}-q_{2\perp}|^2,
\nonumber\\ 
s_{45}\, s_{56} &=& s_{456}\, |p_{5\perp}|^2 = s_{456}\, |q_{2\perp}-q_{1\perp}|^2.
\end{eqnarray}
The mass-shell conditions and \eqn{srel} imply that
\begin{equation}
s_{34}\, s_{45}\, s_{56} = s\, |p_{4\perp}|^2\, |p_{5\perp}|^2. 
\end{equation}
In addition, one can see that
\beq
s_{23} + s_{34} + s_{24} = - |p_{3\perp}+p_{4\perp}|^2 = - |q_{2\perp}|^2.
\eeq
The momentum that flows out along the ladder is $p_4+p_5=q_3-q_1$, with
\beq
|p_{4\perp}+p_{5\perp}|^2 = s_{45} \left(1 - \frac{s\, s_{45}}{s_{345}\, s_{456}} \right)\, 
.\label{eq:mtmcentr}
\eeq

\section{Quasi multi-Regge kinematics}
\label{sec:qmrk}

\subsection{Quasi-multi-Regge kinematics of a pair at either end of the ladder}
\label{sec:nptampqmr}

In the quasi-multi-Regge kinematics of \eqn{qmrknpt}, we require that the gluons
are strongly ordered in rapidity, except for a pair at either end of the ladder.
In light-cone coordinates, it is
\begin{equation}
p_3^+\simeq p_4^+ \gg p_5^+\cdots\gg p_n^+; \qquad p_3^-\simeq p_4^-\ll p_5^- \cdots\ll p_n^-.
\end{equation}
Momentum conservation (\ref{nkin}) then becomes
\begin{eqnarray}
0 &=& \sum_{i=3}^n p_{i\perp}\, ,\nonumber \\
p_2^+ &=& -(p_3^+ + p_4^+)\, ,\label{qmrkin}\\ 
p_1^- &=& -p_n^-\, .\nonumber
\end{eqnarray}
The cyclic Mandelstam invariants are
\bea
s &=& (p_3^+ + p_4^+) p_n^-, \nonumber\\
s_{23} &=& - (p_3^+ + p_4^+) p_3^- = - |p_{3\perp}|^2 - p_4^+p_3^-, \nonumber\\ 
s_{45} &=& p_4^+ p_5^-, \nonumber\\
&\vdots& \nonumber\\
s_{n-1,n} &=& p_{n-1}^+ p_n^-, \nonumber\\
s_{n1} &=& - p_n^+ p_n^- = - |p_{n\perp}|^2 ,
\end{eqnarray}
where we did not indicate $s_{34}$ because it is written as in \eqn{eq:mandelst},
since no approximation is taken on it.

\subsection{Quasi-multi-Regge kinematics of two pairs, one at each end of the ladder}
\label{sec:nptampqmrsq}

In the quasi-multi-Regge kinematics of \eqn{qmrksqnpt}, we require that the gluons
are strongly ordered in rapidity, except for two pairs, one at each end of the ladder.
In light-cone coordinates, it is
\bea
&& p_3^+\simeq p_4^+ \gg p_5^+ \cdots \gg p_{n-2}+\gg p_{n-1}^+ \simeq p_n^+\, ,\nn\\ 
&& p_3^-\simeq p_4^-\ll p_5^- \cdots \ll p_{n-2}-\ll p_{n-1}^- \simeq p_n^-\, .\nn
\eea
Momentum conservation (\ref{nkin}) then becomes
\begin{eqnarray}
0 &=& \sum_{i=3}^n p_{i\perp}\, ,\nonumber \\
p_2^+ &=& -(p_3^+ + p_4^+)\, ,\label{qmrsqkin}\\ 
p_1^- &=& -(p_{n-1}^- + p_n^-)\, .\nonumber
\end{eqnarray}
The cyclic Mandelstam invariants are
\bea
s &=& (p_3^+ + p_4^+) (p_{n-1}^- + p_n^-), \nonumber\\
s_{23} &=& - (p_3^+ + p_4^+) p_3^- = - |p_{3\perp}|^2 - p_4^+p_3^-, \nonumber\\ 
s_{45} &=& p_4^+ p_5^-, \nonumber\\
&\vdots& \nonumber\\
s_{n-2,n-1} &=& p_{n-2}^+ p_{n-1}^-, \nonumber\\
s_{1n} &=& - p_n^+ (p_{n-1}^- + p_n^-) = - |p_{n\perp}|^2 - p_n^+ p_{n-1}^-,
\end{eqnarray}
where we did not indicate $s_{34}$ and $s_{n-1,n}$ because 
no approximation is taken on them. It is easy to see that
\bea
s_{234} &=& - |p_{3\perp} + p_{4\perp}|^2, \nn\\
s_{n-1,n,1} &=& - |p_{n-1\perp} + p_{n\perp}|^2.
\eea

\subsection{Quasi-multi-Regge kinematics of a pair along the ladder}
\label{sec:nptampqmrc}

We require that the gluons
are strongly ordered in rapidity, except for a pair along the ladder.
In light-cone coordinates, it is
\begin{equation}
p_3^+\gg p_4^+ \simeq p_5^+\gg p_6^+; \qquad p_3^-\ll p_4^-\simeq p_5^- \ll p_6^-.
\end{equation}
Momentum conservation (\ref{nkin}) then becomes
\begin{eqnarray}
0 &=&  p_{3\perp} + p_{4\perp} + p_{5\perp} + p_{6\perp}\, ,\nonumber \\
p_2^+ &=& -p_3^+\, ,\label{qmrkinc}\\ 
p_1^- &=& -p_6^-\, .\nonumber
\end{eqnarray}
The cyclic Mandelstam invariants are
\bea
s &=& p_3^+ p_6^-, \nonumber\\
s_{23} &=& - p_3^+ p_3^- = - |p_{3\perp}|^2, \nonumber\\ 
s_{34} &=& p_3^+ p_4^-, \nonumber\\
s_{56} &=& p_5^+ p_6^-, \nonumber\\
s_{61} &=& - p_6^+ p_6^- = - |p_{6\perp}|^2\, ,\label{centrkin}
\end{eqnarray}
where we did not indicate $s_{45}$ since no approximation is taken on it.
The mass-shell conditions for the gluons emitted along the ladder are
\beq
|p_{4\perp}|^2 = \frac{s_{34}s_{46}}{s}, \qquad |p_{5\perp}|^2 = \frac{s_{35}s_{56}}{s}\, .
\eeq
In addition, it is useful to evaluate
\beq
s_{234} = - |p_{3\perp} + p_{4\perp}|^2 - p_4^-p_5^+\, .
\eeq

\subsection{Quasi-multi-Regge kinematics of three-of-a-kind}
\label{sec:kin6qmr3}

In the quasi-multi-Regge kinematics of \sec{sec:nptampqmrc}, where the outgoing
gluons are emitted three in a cluster on one end and one on the other end of the ladder,
\begin{equation}
p_3^+\simeq p_4^+ \simeq p_5^+\gg p_6^+; \qquad p_3^-\simeq p_4^-\simeq p_5^- \ll p_6^-\, .
\end{equation}
Momentum conservation (\ref{nkin}) then becomes
\begin{eqnarray}
0 &=&  p_{3\perp} + p_{4\perp} + p_{5\perp} + p_{6\perp}\, ,\nonumber \\
p_2^+ &=& - (p_3^+ + p_4^+ + p_5^+)\, ,\label{qmrkin3g}\\ 
p_1^- &=& -p_6^-\, .\nonumber
\end{eqnarray}
The cyclic Mandelstam invariants are
\bea
s &=& p_3^+ p_6^-\, , \nonumber\\
s_{23} &=& - (p_3^+ + p_4^+ + p_5^+) p_3^- = - |p_{3\perp}|^2 - (p_4^+ + p_5^+) p_3^-\, ,\nonumber\\ 
s_{56} &=& p_5^+ p_6^-\, , \nonumber\\
s_{61} &=& - p_6^+ p_6^- = - |p_{6\perp}|^2\, ,\label{kin3g}
\end{eqnarray}
where we did not indicate $s_{34}$ and $s_{45}$ since no approximation is taken on them.
In addition, it is useful to evaluate
\beq
s_{234} = - |p_{3\perp} + p_{4\perp}|^2 - (p_3^- + p_4^-) p_5^+\, .
\eeq


\begin{thebibliography}{999}

\bibitem{Bern:2005iz}
  Z.~Bern, L.~J.~Dixon and V.~A.~Smirnov,
  ``Iteration of planar amplitudes in maximally supersymmetric Yang-Mills
  theory at three loops and beyond,''
  Phys.\ Rev.\  D {\bf 72} (2005) 085001
  [arXiv:hep-th/0505205].
  
\bibitem{Bern:2006vw}
  Z.~Bern, M.~Czakon, D.~A.~Kosower, R.~Roiban and V.~A.~Smirnov,
  ``Two-loop iteration of five-point N = 4 super-Yang-Mills amplitudes,''
  Phys.\ Rev.\ Lett.\  {\bf 97} (2006) 181601
  [arXiv:hep-th/0604074].
  
\bibitem{Cachazo:2008vp}
F.~Cachazo, M.~Spradlin and A.~Volovich,
``Iterative structure within the five-particle two-loop amplitude,''
Phys.\ Rev.\  D {\bf 74}, 045020 (2006)
[arXiv:hep-th/0602228].

  
\bibitem{Korchemsky:1987wg}
  G.~P.~Korchemsky and A.~V.~Radyushkin,
  ``Renormalization of the Wilson Loops Beyond the Leading Order,''
  Nucl.\ Phys.\  B {\bf 283} (1987) 342.
    
\bibitem{Beisert:2006ez}
  N.~Beisert, B.~Eden and M.~Staudacher,
  ``Transcendentality and crossing,''
  J.\ Stat.\ Mech.\  {\bf 0701} (2007) P021
  [arXiv:hep-th/0610251].
    
\bibitem{Magnea:1990zb}
  L.~Magnea and G.~Sterman,
  ``Analytic continuation of the Sudakov form-factor in QCD,''
  Phys.\ Rev.\  D {\bf 42} (1990) 4222.

\bibitem{Sterman:2002qn}
  G.~Sterman and M.~E.~Tejeda-Yeomans,
  ``Multi-loop amplitudes and resummation,''
  Phys.\ Lett.\  B {\bf 552} (2003) 48
  [arXiv:hep-ph/0210130].
  
  \bibitem{Alday:2007he}
  L.~F.~Alday and J.~Maldacena,
  ``Comments on gluon scattering amplitudes via AdS/CFT,''
  JHEP {\bf 0711} (2007) 068
  [arXiv:0710.1060 [hep-th]].

\bibitem{Alday:2007hr}
  L.~F.~Alday and J.~M.~Maldacena,
  ``Gluon scattering amplitudes at strong coupling,''
  JHEP {\bf 0706} (2007) 064
  [arXiv:0705.0303 [hep-th]].
  
\bibitem{Drummond:2007bm}
  J.~M.~Drummond, J.~Henn, G.~P.~Korchemsky and E.~Sokatchev,
  ``The hexagon Wilson loop and the BDS ansatz for the six-gluon amplitude,''
  arXiv:0712.4138 [hep-th].

\bibitem{Bern:2008ap}
  Z.~Bern, L.~J.~Dixon, D.~A.~Kosower, R.~Roiban, M.~Spradlin, C.~Vergu and A.~Volovich,
  ``The Two-Loop Six-Gluon MHV Amplitude in Maximally Supersymmetric Yang-Mills
  Theory,''
  arXiv:0803.1465 [hep-th].

\bibitem{Cachazo:2008hp}
  F.~Cachazo, M.~Spradlin and A.~Volovich,
  ``Leading Singularities of the Two-Loop Six-Particle MHV Amplitude,''
  arXiv:0805.4832 [hep-th].

\bibitem{Drummond:2008aq}
  J.~M.~Drummond, J.~Henn, G.~P.~Korchemsky and E.~Sokatchev,
  ``Hexagon Wilson loop = six-gluon MHV amplitude,''
  arXiv:0803.1466 [hep-th].
  
\bibitem{Kuraev:1976ge}
  E.~A.~Kuraev, L.~N.~Lipatov and V.~S.~Fadin,
  ``Multi - Reggeon Processes In The Yang-Mills Theory,''
  Sov.\ Phys.\ JETP {\bf 44} (1976) 443
  [Zh.\ Eksp.\ Teor.\ Fiz.\  {\bf 71} (1976) 840].
  
\bibitem{Bartels:2008ce}
  J.~Bartels, L.~N.~Lipatov and A.~Sabio Vera,
  ``BFKL Pomeron, Reggeized gluons and Bern-Dixon-Smirnov amplitudes,''
  Phys.\ Rev.\  D {\bf 80} (2009) 045002
  [arXiv:0802.2065 [hep-th]].
  
\bibitem{Schabinger:2009bb}
  R.~M.~Schabinger,
  ``The Imaginary Part of the N = 4 Super-Yang-Mills Two-Loop Six-Point MHV
  Amplitude in Multi-Regge Kinematics,''
  arXiv:0910.3933 [hep-th].


\bibitem{Mangano:1990by}
  M.~L.~Mangano and S.~J.~Parke,
  ``Multiparton amplitudes in gauge theories,''
  Phys.\ Rept.\  {\bf 200} (1991) 301
  [arXiv:hep-th/0509223].

\bibitem{DelDuca:1993pp}
  V.~Del Duca,
  ``Parke-Taylor amplitudes in the multi - Regge kinematics,''
  Phys.\ Rev.\  D {\bf 48} (1993) 5133
  [arXiv:hep-ph/9304259].

\bibitem{DelDuca:1995zy}
V.~Del Duca,
``Equivalence of the Parke-Taylor and the Fadin-Kuraev-Lipatov amplitudes 
in the high-energy limit,''
Phys.\ Rev.\  D {\bf 52} (1995) 1527 [arXiv:hep-ph/9503340].

\bibitem{DelDuca:1999rs}
  V.~Del Duca, L.~J.~Dixon and F.~Maltoni,
  ``New color decompositions for gauge amplitudes at tree and loop level,''
  Nucl.\ Phys.\  B {\bf 571} (2000) 51
  [arXiv:hep-ph/9910563].

\bibitem{DelDuca:1996km}
V.~Del Duca, 
``Next-to-leading corrections to the BFKL equation,''
[arXiv:hep-ph/9605404].

\bibitem{Lipatov:1976zz}
  L.~N.~Lipatov,
  ``Reggeization Of The Vector Meson And The Vacuum Singularity In Nonabelian
  Gauge Theories,''
  Sov.\ J.\ Nucl.\ Phys.\  {\bf 23} (1976) 338
  [Yad.\ Fiz.\  {\bf 23} (1976) 642].

\bibitem{Lipatov:1991nf}
  L.~N.~Lipatov,
  ``High-energy scattering in QCD and in quantum gravity and two-dimensional
  field theories,''
  Nucl.\ Phys.\  B {\bf 365} (1991) 614.

\bibitem{Fadin:1993wh}
  V.~S.~Fadin and L.~N.~Lipatov,
  ``Radiative corrections to QCD scattering amplitudes in a multi - Regge
  kinematics,''
  Nucl.\ Phys.\  B {\bf 406} (1993) 259.

\bibitem{DelDuca:1998kx}
  V.~Del Duca and C.~R.~Schmidt,
  ``Virtual next-to-leading corrections to the impact factors in the
  high-energy limit,''
  Phys.\ Rev.\  D {\bf 57} (1998) 4069
  [arXiv:hep-ph/9711309].
  
\bibitem{DelDuca:1998cx}
  V.~Del Duca and C.~R.~Schmidt,
  ``Virtual next-to-leading corrections to the Lipatov vertex,''
  Phys.\ Rev.\  D {\bf 59} (1999) 074004
  [arXiv:hep-ph/9810215].

\bibitem{DelDuca:2008pj}
  V.~Del Duca and E.~W.~N.~Glover,
  ``Testing high-energy factorization beyond the next-to-leading-logarithmic
  accuracy,''
  JHEP {\bf 0805} (2008) 056
  [arXiv:0802.4445 [hep-th]].

\bibitem{Bern:1998sc}
  Z.~Bern, V.~Del Duca and C.~R.~Schmidt,
  ``The infrared behavior of one-loop gluon amplitudes at
  next-to-next-to-leading order,''
  Phys.\ Lett.\  B {\bf 445} (1998) 168
  [arXiv:hep-ph/9810409].

\bibitem{Fadin:1992zt}
  V.~S.~Fadin and R.~Fiore,
  ``Quark Contribution To The Gluon-Gluon - Reggeon Vertex In QCD,''
  Phys.\ Lett.\  B {\bf 294} (1992) 286.

\bibitem{Fadin:1993qb}
  V.~S.~Fadin, R.~Fiore and A.~Quartarolo,
  ``Radiative corrections to quark quark reggeon vertex in QCD,''
  Phys.\ Rev.\  D {\bf 50} (1994) 2265
  [arXiv:hep-ph/9310252].
  
\bibitem{Fadin:1995xg}
  V.~S.~Fadin, M.~I.~Kotsky and R.~Fiore,
  ``Gluon Reggeization In QCD In The Next-To-Leading Order,''
  Phys.\ Lett.\  B {\bf 359} (1995) 181.

\bibitem{Fadin:1995km}
  V.~S.~Fadin, R.~Fiore and A.~Quartarolo,
  ``Reggeization of quark quark scattering amplitude in QCD,''
  Phys.\ Rev.\  D {\bf 53} (1996) 2729
  [arXiv:hep-ph/9506432].

\bibitem{Fadin:1996tb}
  V.~S.~Fadin, R.~Fiore and M.~I.~Kotsky,
  ``Gluon Regge trajectory in the two-loop approximation,''
  Phys.\ Lett.\  B {\bf 387} (1996) 593
  [arXiv:hep-ph/9605357].

\bibitem{Blumlein:1998ib}
  J.~Bl\"umlein, V.~Ravindran and W.~L.~van Neerven,
  ``On the gluon Regge trajectory in ${\cal O}(\alpha_s^2$),''
  Phys.\ Rev.\  D {\bf 58} (1998) 091502
  [arXiv:hep-ph/9806357].

\bibitem{DelDuca:2001gu}
  V.~Del Duca and E.~W.~N.~Glover,
  ``The high energy limit of QCD at two loops,''
  JHEP {\bf 0110} (2001) 035
  [arXiv:hep-ph/0109028].

\bibitem{Kotikov:2000pm}
  A.~V.~Kotikov and L.~N.~Lipatov,
  ``NLO corrections to the BFKL equation in QCD and in supersymmetric gauge
  theories,''
  Nucl.\ Phys.\  B {\bf 582} (2000) 19
  [arXiv:hep-ph/0004008].
  
\bibitem{Kotikov:2002ab}
  A.~V.~Kotikov and L.~N.~Lipatov,
  ``DGLAP and BFKL equations in the N=4 supersymmetric gauge theory,''
  Nucl.\ Phys.\  B {\bf 661} (2003) 19
  [Erratum-ibid.\  B {\bf 685} (2004) 405]
  [arXiv:hep-ph/0208220].

\bibitem{Drummond:2007aua}
  J.~M.~Drummond, G.~P.~Korchemsky and E.~Sokatchev,
  ``Conformal properties of four-gluon planar amplitudes and Wilson loops,''
  Nucl.\ Phys.\  B {\bf 795} (2008) 385
  [arXiv:0707.0243 [hep-th]].

\bibitem{Naculich:2007ub}
  S.~G.~Naculich and H.~J.~Schnitzer,
  ``Regge behavior of gluon scattering amplitudes in N=4 SYM theory,''
  Nucl.\ Phys.\  B {\bf 794} (2008) 189
  [arXiv:0708.3069 [hep-th]].

\bibitem{Bern:2006ew}
  Z.~Bern, M.~Czakon, L.~J.~Dixon, D.~A.~Kosower and V.~A.~Smirnov,
  ``The Four-Loop Planar Amplitude and Cusp Anomalous Dimension in Maximally
  Supersymmetric Yang-Mills Theory,''
  Phys.\ Rev.\  D {\bf 75} (2007) 085010
  [arXiv:hep-th/0610248].

\bibitem{Cachazo:2006az}
  F.~Cachazo, M.~Spradlin and A.~Volovich,
  ``Four-Loop Cusp Anomalous Dimension From Obstructions,''
  Phys.\ Rev.\  D {\bf 75} (2007) 105011
  [arXiv:hep-th/0612309].
  
\bibitem{Spradlin:2008uu}
  M.~Spradlin, A.~Volovich and C.~Wen,
  ``Three-Loop Leading Singularities and BDS Ansatz for Five Particles,''
  arXiv:0808.1054 [hep-th].
  
\bibitem{Anastasiou:2003kj}
  C.~Anastasiou, Z.~Bern, L.~J.~Dixon and D.~A.~Kosower,
  ``Planar amplitudes in maximally supersymmetric Yang-Mills theory,''
  Phys.\ Rev.\ Lett.\  {\bf 91} (2003) 251602
  [arXiv:hep-th/0309040].

\bibitem{us}
  V.~Del Duca, C.~Duhr and E.~W.~Nigel Glover,
  ``The five-gluon amplitude in the high-energy limit,''
  arXiv:0905.0100 [hep-th].
 
\bibitem{Drummond:2007au}
  J.~M.~Drummond, J.~Henn, G.~P.~Korchemsky and E.~Sokatchev,
  ``Conformal Ward identities for Wilson loops and a test of the duality with
  gluon amplitudes,''
  arXiv:0712.1223 [hep-th].

\bibitem{Brower:2008nm}
  R.~C.~Brower, H.~Nastase, H.~J.~Schnitzer and C.~I.~Tan,
  ``Implications of multi-Regge limits for the Bern-Dixon-Smirnov conjecture,''
  Nucl.\ Phys.\  B {\bf 814} (2009) 293
  [arXiv:0801.3891 [hep-th]].
  
\bibitem{Brower:2008ia}
  R.~C.~Brower, H.~Nastase, H.~J.~Schnitzer and C.~I.~Tan,
  ``Analyticity for Multi-Regge Limits of the Bern-Dixon-Smirnov Amplitudes,''
  Nucl.\ Phys.\  B {\bf 822} (2009) 301
  [arXiv:0809.1632 [hep-th]].

\bibitem{DelDuca:1995ki}
  V.~Del Duca,
  ``Real next-to-leading corrections to the multigluon amplitudes in the
  helicity formalism,''
  Phys.\ Rev.\  D {\bf 54} (1996) 989
  [arXiv:hep-ph/9601211].


\bibitem{Fadin:1996nw}
  V.~S.~Fadin and L.~N.~Lipatov,
  ``Next-to-leading Corrections to the BFKL Equation From the Gluon and Quark
  Production,''
  Nucl.\ Phys.\  B {\bf 477} (1996) 767
  [arXiv:hep-ph/9602287].


\bibitem{Del Duca:1999ha}
  V.~Del Duca, A.~Frizzo and F.~Maltoni,
  ``Factorization of tree QCD amplitudes in the high-energy limit and in  the
  collinear limit,''
  Nucl.\ Phys.\  B {\bf 568} (2000) 211
  [arXiv:hep-ph/9909464].
  
  
 

    
\end{thebibliography}
\end{document}